\documentclass[superscriptaddress,amsmath,amssymb,aps,prl,reprint]{revtex4-1}
\usepackage{graphicx}
\usepackage{dcolumn}
\usepackage{bm}
\usepackage{amssymb}
\usepackage{amsmath}
\usepackage{notes2bib}
\usepackage{siunitx}
\usepackage{xcolor}
\usepackage[utf8]{inputenc}
\usepackage{upgreek}
\usepackage{mhchem}

\usepackage{hyperref}

\begin{document}

\title{Proximity-induced superconductivity in a bilayer graphene quantum point contact}

\author{Clara Galante-Agero}
\email{cgalante@phys.ethz.ch}
\author{Christoph Adam}
\author{Artem O. Denisov}
\author{Jonas D. Gerber}
\author{Markus Niese}
\author{Alexandra Mestre-Tor\`a}
\author{Marta Perego}
\author{Jessica Richter}

\affiliation{Laboratory for Solid State Physics, ETH Zurich,~CH-8093~Zurich, Switzerland}
\author{Takashi Taniguchi}
\affiliation{Research Center for Materials Nanoarchitectonics, National Institute for Materials Science,  1-1 Namiki, Tsukuba 305-0044, Japan}
\author{Kenji Watanabe}
\affiliation{Research Center for Electronic and Optical Materials, National Institute for Materials Science, 1-1 Namiki, Tsukuba 305-0044, Japan}

\author{Klaus Ensslin}
\author{Thomas Ihn}

\affiliation{Laboratory for Solid State Physics, ETH Zurich,~CH-8093~Zurich, Switzerland}
\affiliation{Quantum Center, ETH Zurich,~CH-8093 Zurich, Switzerland}

\date{\today}

\begin{abstract}
{
We report the realization of a gate-defined quantum point contact (QPC) in bilayer graphene proximitized by a single aluminum superconducting electrode. Superconducting correlations induced in the ballistic channel enhance the conductance plateaus beyond their normal-state values. In addition, we observe a pronounced above-gap conductance anomaly which serves as a spectroscopic signature of the loss of superconductivity and the associated collapse of the Andreev excess current. By reconstructing the nonlinear current-voltage characteristics, we find that the magnitude of the excess current increases as successive QPC modes are populated. 
Additionally, we find that the switching current associated with the loss of superconductivity follows the underlying mode structure of the QPC, exhibiting discrete levels consistent with a heat dissipation-driven transition. These results demonstrate that the one-dimensional transport modes of the QPC govern both the equilibrium proximity effect and the non-equilibrium dynamics of the hybrid system.}
\end{abstract}

\maketitle

The integration of superconductivity with ballistic semiconductor devices offers a pathway to manipulate quantum states at the interface of normal and superconducting matter. Bilayer graphene (BLG), with its electrically tunable band gap \cite{oostinga2008gate}, provides a versatile platform for such hybrid architectures, hosting gate-defined ballistic nanostructures including quantum point contacts (QPCs) \cite{QPC_hiske,PhysRevLett.124.177701,lee2020tunable}, quantum dots \cite{eich2018spin,Banszerus18,Eich18,PhysRevLett.123.026803},
topological valley splitters \cite{valley_valve,valley_polarizer},
and interferometric devices \cite{Fabry_perot, AB_oscillations, Deprez2021}.

When coupled to a superconductor, few-layer graphene exhibits the proximity effect \cite{PhysRevLett.97.067007,bipolar_supercurrent_Pablo,Sato07,PhysRevB.77.184507,PhysRevB.79.165436, Choi13}, enabling hybrid devices in which both the electronic band structure and superconducting correlations can be controlled electrostatically \cite{Sato07,lee2018proximity}.
While graphene Josephson junctions have been extensively studied \cite{bipolar_supercurrent_Pablo, Calado2015, Quantum_oscillations_JJ, S_and_quantum_hall, Tayloring_supercurrents, ABS_resonator}, superconducting transport through gate-defined one-dimensional ballistic channels in bilayer graphene 
remains unexplored.

At a superconducting--normal (SN) interface, subgap transport is governed by Andreev reflection, in which an incoming electron from the normal metal is reflected back as a hole while a Cooper pair enters the superconductor \cite{blonder1982transition,beenakker1992quantum}. For transparent interfaces, Andreev processes enhance the conductance beyond its normal-state value and generate a finite excess current at voltages above the superconducting gap \cite{blonder1982transition, BTK_cunb_pointcontacts, Zaitsev1980, Artemenko1979, octavio1983subharmonic}. 
At higher bias, the Andreev-induced excess current gradually reduces due to nonequilibrium dissipation, which eventually leads to a breakdown of superconductivity. The excess current is well understood in standard superconducting junctions. However, it remains unexplored experimentally how ballistic QPC modes affect superconducting transport in this regime, and whether they can serve as a unifying parameter for both equilibrium Andreev processes and nonequilibrium superconducting breakdown.

Here, we address these questions experimentally in a single device. We realize a gate-defined bilayer graphene quantum point contact proximitized by a single aluminum superconducting electrode. We observe conductance plateaus enhanced beyond their normal-state quantization due to Andreev reflections at the superconducting interface. Finite-bias spectroscopy reveals both the superconducting gap and a pronounced above-gap conductance anomaly associated with the suppression of superconductivity and thereby of the excess current. We extract the excess current from the nonlinear transport characteristics and find that it evolves systematically as successive QPC modes are populated. Moreover, in current-biased measurements, the switching current associated with the collapse of superconductivity also evolves in discrete steps as individual transport modes are opened. Together, these observations demonstrate that the quantized one-dimensional modes of the QPC govern both the equilibrium Andreev transport and the nonequilibrium breakdown of superconductivity in the hybrid device.

\begin{figure}[t]
    \centering
    \includegraphics[width=1.0\linewidth]{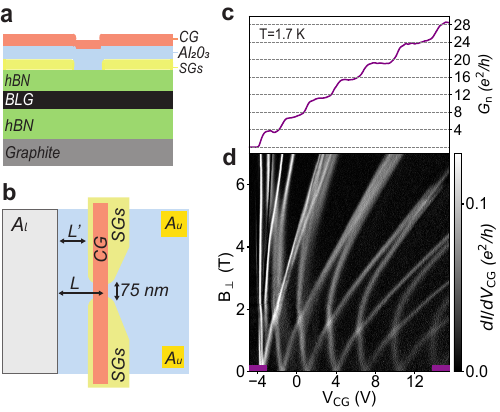}
    \caption{(a) Schematic cross section of the device. The bilayer graphene (BLG) encapsulated in hexagonal boron nitride (hBN), and the channel gate (CG) and split gates (SGs) are indicated. (b) Schematic top view of the device with aluminium $Al$ and gold $Au$ contacts to the graphene displayed. The distances $L$ and $L'$ are approximately $\SI{350}{nm}$ and $\SI{100}{nm}$, respectively. The gap between SGs is $\SI{75}{nm}$ as indicated. (c) Conductance measurements at $\SI{1.7}{K}$ and $B_\perp=0$. $G_\textnormal{n}$ denotes the QPC conductance in the normal state. (d) Transconductance of the QPC as a function of the out-of-plane magnetic field and CG voltage $V_\textnormal{CG}$. Black regions correspond to plateaus in conductance.  }
    \label{Fig1}
\end{figure}

Figure \ref{Fig1}(a) depicts a cross-sectional schematic of the device, consisting of a BLG flake encapsulated in hexagonal boron nitride, a bottom graphite gate, and two layers of top metallic gates separated by an \ce{Al2O3} dielectric. Figure \ref{Fig1}(b) illustrates the top-view design of the hybrid QPC device. The two split gates (SGs, yellow), separated by a lithographic gap of $\SI{75}{nm}$, are used together with the graphite back gate to open a band gap in the BLG and electrostatically confine charge carriers into a one-dimensional channel. A channel gate (CG), vertically separated from the split gates by an \ce{Al2O3} layer, is used to control the carrier density in the constriction and thereby the mode occupation of the QPC, similar to Refs.~\cite{QPC_hiske, PhysRevLett.124.177701, Jonas_QPC}. The QPC is coupled to a superconducting aluminum contact (Ti/Al, 5/40 nm) located at $L \approx \SI{350}{nm}$ from the narrowest point of the constriction.

We first investigate the QPC properties at 1.7 K, which is above the critical temperature of the superconducting contact, ensuring that it remains in the normal state. Fig.~\ref{Fig1}(c) displays the linear conductance at zero out-of-plane magnetic field as a function of the channel gate voltage $V_\textnormal{CG}$. We observe conductance plateaus at values close to $G_{n}=(4e^2/h)N$ ($N=1,2,3,\ldots$), where the factor of 4 reflects the spin and valley degeneracy of BLG \cite{topologically_states_QPC}. We use $G_{n}$ to denote the QPC conductance in the normal state. The ballistic nature of the channel is further supported by the transconductance map measured under an out-of-plane magnetic field $B_\perp$ (Fig.~\ref{Fig1}(d)): as $B_\perp$ increases, the four-fold degenerate modes split due to the valley Zeeman effect \cite{lee2020tunable} and smoothly evolve into quantum Hall edge states at high fields. The persistence of these features down to zero magnetic field indicates that the QPC operates in the ballistic regime despite the proximity of the metallic contact.
We note that full pinch-off is not achieved in this device, and a residual conductance corresponding to a parallel resistance of approximately $R_\parallel= \SI{5}{k\Omega}$ remains. We attribute this to parallel conduction paths under the split-gated region. Such incomplete pinch-off can happen in gate-defined bilayer graphene devices, where the split gates may not fully suppress current flow.
Accordingly, all conductance traces are corrected for series resistance and parallel conductance, as detailed in the Supplementary material.


We now investigate the system at $\SI{10}{mK}$, below the critical temperature of Al, and discuss how its superconductivity modifies transport through the QPC. We first examine the conductance enhancement induced by Andreev reflection and then use finite-bias spectroscopy to characterize both the superconducting gap and the above-gap anomaly associated with the loss of superconductivity.

\begin{figure*}[t!]
    \centering
    \includegraphics[width=1\linewidth]{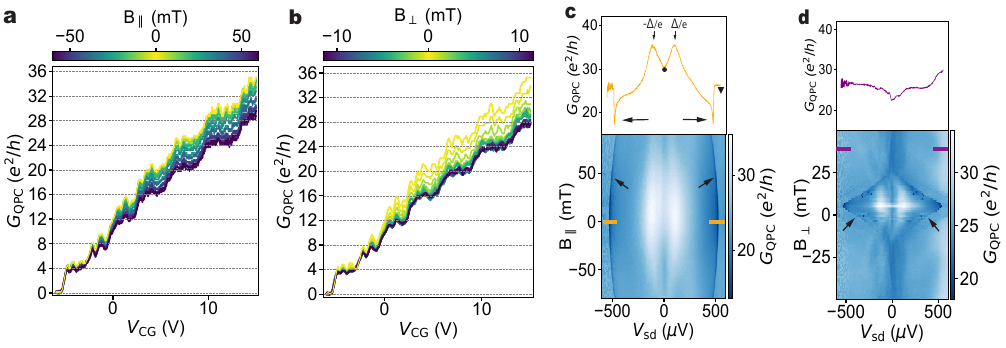}
    \caption{Conductance measurements at \SI{10}{mK}. (a,b) QPC conductance as a function of channel gate voltage $V_\textnormal{CG}$, and in-plane $B_\parallel$ (a) and out-of-plane $B_\perp$ (b) magnetic field. (c) $ G_\textnormal{QPC}$ as a function of in-plane field $B_\parallel$ and source-drain bias voltage $V_\textnormal{sd}$, at a fixed channel gate voltage $V_\textnormal{CG}=\SI{11}{V}$. The top curve shows the conductance trace at  $B_\parallel=0$. Two conductance peaks are observed, from which the superconducting gap, $\Delta$, is extracted. Black arrows indicate the anomalous dip in conductance, marking the collapse of the Andreev excess current. (d) Conductance as a function of out-of-plane field $B_\perp$ and bias voltage $V_\textnormal{sd}$, at the same fixed channel gate voltage $V_\textnormal{CG}=\SI{11}{V}$. The top curve shows the conductance trace at $B_\perp=\SI{40}{mT}$, therefore higher than the critical magnetic field $B_{c,\perp}\sim\SI{20}{mT}$, where no double peak and anomalous dips are observed.}
    \label{Fig2}
\end{figure*}

Figures~\ref{Fig2}(a,b) show the linear QPC conductance $G_\textnormal{QPC}$ as a function of channel gate voltage for varying in-plane $B_\parallel$ and out-of-plane $B_\perp$ magnetic fields, respectively. 
At $B_{\perp,\parallel}=0$ (yellow traces), where the contact is superconducting, $G_\textnormal{QPC}$ exhibits a step-like evolution with plateau height that systematically exceeds the normal-state quantization $G_n=(4e^2/h)N$. Specifically, the conductance increases in steps of approximately $1.25 \times 4e^2/h$. We interpret this enhancement as the manifestation of Andreev reflections at the SN interface. Within the Blonder-Tinkham-Klapwijk (BTK) framework, 
a perfectly transparent interface would yield a conductance doubling, $2\times G_n$.
Our observed enhancement factor $G_\textnormal{s}/G_\textnormal{n}\approx 1.25$ 
corresponds to
an effective interface transparency of $\tau \approx 0.9$ (see Supplementary).
This deviation from the ideal limit $(\tau =1)$ may arise from residual interface scattering or Fermi-wave-vector mismatch \cite{blonder1982transition,FWM}.

The evolution of the conductance traces, $G_\textnormal{QPC}(B_{\parallel,\perp})$, as a function of magnetic field in Figs.~\ref{Fig2}(a,b) captures the transition from the superconducting to the normal state; as $B$ increases (moving toward the darker traces in Figs.~\ref{Fig2}(a,b)), the Andreev enhancement is gradually suppressed, and the conductance plateaus return to their normal-state values, in agreement with previous studies in semiconductor quantum wells and nanowires \cite{hybrid_arrays_2024,nanowires_2025}. As expected for thin-film superconductors, the Andreev enhancement in $G_\textnormal{QPC}$ persists up to in-plane magnetic fields $B_\parallel$ (Fig. \ref{Fig2}(a)) an order of magnitude larger than in the out-of-plane $B_\perp$ configuration (Fig. \ref{Fig2}(b)), where orbital effects strongly suppress superconductivity. Within the narrow field range where superconductivity prevails, neither valley nor spin splitting of the 1D modes is resolved. Reproducible mesoscopic fluctuations remain superimposed on the conductance plateaus, which we attribute to partial reflections between the QPC and the superconducting contact.

Having established that the enhanced conductance originates from Andreev reflection, we next investigate its finite-bias response. We perform finite-bias spectroscopy by measuring the differential conductance $G_\textnormal{QPC}$ as a function of d.c.\ source-drain bias voltage $V_\textnormal{sd}$, at the fixed channel-gate voltage $V_\textnormal{CG}=\SI{11}{V}$, on the $N=6$ conductance plateau. The top curves in Figs.~\ref{Fig2}(c,d) show the results for $B_{\parallel}=0$ (yellow) and $B_\perp=\SI{40}{mT}$ (purple), where the latter is sufficient to suppress superconductivity in the contact. The zero in-plane field curve (top in Fig.~\ref{Fig2}(c)) exhibits the characteristic BTK line shape of a partially transparent SN interface, including enhanced zero-bias conductance (black circle) relative to the normal state value (black triangle) and coherence peaks at $|V_\textnormal{sd}|=\Delta/e$. From the peak positions we extract the superconducting energy gap of the aluminum contact $\Delta \approx 120~\mu$V.

For the \SI{40}{nm} thick Al film used in our device, the superconducting gap is expected to be close to its bulk value of \SI{180}{\micro V} \cite{Gap_vs_d_spectroscopy, Bulk_value}. However, previous studies have shown that Ti/Al or Ti/Al/Au heterostructures exhibit a reduced critical temperature and accordingly a smaller induced superconducting gap \cite{MARTINIS200023, ALO_junctions, YongJoo_2010}, which is consistent with the reduced gap observed here. The rounded shape of the gap features, compared with the sharper structure expected from the BTK model, may originate from a reentrance effect commonly associated with diffusive SN junctions~\cite{Jehl2000,PhysRevLett.84.3398}. In our device, the finite size of the ballistic cavity between the QPC and the superconducting region can introduce an additional Thouless energy scale. When this scale becomes comparable to the gap, the corresponding features are expected to be broadened or even shifted~\cite{Courtois1999}. 

Beyond the gap-related features, the differential conductance in Fig.~\ref{Fig2}(c) exhibits an anomalous dip at bias voltages significantly larger than the gap, indicated by the horizontal arrows in the yellow trace of Fig.~\ref{Fig2}(c). This feature plays a central role in the remainder of this work.
Such dips have been observed both in SN junctions \cite{Xiong_SN_junction, jiang2016unconventional, WESTBROOK_1999, Gifford_2016} and Josephson junctions \cite{anomalus_MAR, MAR_gJJ, YongJoo_2010, Tomi_2021} and are commonly attributed to the suppression of the excess current resulting from the transition of the contact from the superconducting into the normal state \cite{Tomi_2021}.
When superconductivity in the contact is suppressed by applying an out-of-plane field (see purple curve in Fig.~\ref{Fig2}(c)), both the gap-related double-peak structure and the above-gap anomaly disappear.

The color plots in  Figs.~\ref{Fig2}(c, d) show the evolution of the explained features in $B_\perp$ and $B_\parallel$ respectively, at the same voltage configuration ($V_\textnormal{CG}=\SI{11}{V}$). The dark outer features in these maps, indicated by the black arrows, correspond to the anomalous dip, i.e., to the nonequilibrium collapse of superconductivity in the contact, and therefore trace the boundary of the superconducting regime.
The extracted critical fields are determined by fitting the switching current to a linear decay $I_\textnormal{sw}(B_\perp)=I_\textnormal{sw}(0)(1-B_\perp/B_{\textnormal{c},\perp})$ for out-of-plane fields and a quadratic Ginzburg--Landau relation $I_\textnormal{sw}(B_\parallel)=I_\textnormal{sw}(0)[1-(B_\parallel/B_{\textnormal{c},\parallel})^2]$ for in-plane fields.
These fits yield $B_{c,\perp}\approx\SI{20}{mT}$ and $B_{c,\parallel}\approx\SI{320}{mT}$ (see Supplementary), values that are consistent with measurements on independent Ti/Al Hall-bar structures. They confirm that the observed conductance anomaly is a direct signature of the contact's superconducting state.


We now turn to the origin of the pronounced conductance dip observed at biases above the superconducting gap.
In a transparent superconducting–normal junction, the current at biases above the gap can be phenomenologically written as \cite{blonder1982transition, BTK_cunb_pointcontacts}
\begin{equation}
I=V_\textnormal{sd}/R+I_\textnormal{exc}, 
\label{IV prediction}
\end{equation}
where $R$ is the normal-state resistance of the device and $I_\textnormal{exc}$ is the excess current arising from Andreev reflections, which is directly proportional to the superconducting gap. This contribution appears as a current offset in the current–voltage characteristics at high biases $|V_\textnormal{sd}|>\Delta/e$.

When the superconducting gap closes, the excess current vanishes and transport crosses over to a purely dissipative current, giving rise to the conductance dip discussed above. Several mechanisms have been proposed to describe the superconducting-to-normal transition, including pair-breaking current density \cite{Xiong_SN_junction, jiang2016unconventional}, anomalous Andreev processes \cite{anomalus_MAR}, and magnetic stray fields \cite{WESTBROOK_1999, Gifford_2016}
Recently, the superconducting transition has been attributed to an enhanced electron temperature due to Joule heating dissipated in the contact near the junction \cite{YongJoo_2010, Tomi_2021, Joule_spectroscopy,  PhysRevB.109.L140501}. Because typical cooling mechanisms, such as quasiparticle diffusion into the leads and electron--phonon coupling become inefficient at low temperatures, the electron temperature increases locally, provoking the transition to the normal state once the critical temperature is reached. This occurs at a critical dissipated power $P_\textnormal{c} = I_\textnormal{sw}^2 R$, where $I_\textnormal{sw}$ is the switching current and $R$ is the device resistance.

\begin{figure}[t]
    \centering
    \includegraphics[width=1\linewidth]{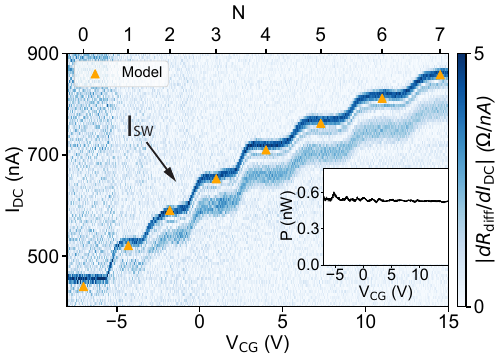}
    \caption{ Switching current as a function of channel gate, when the device is tuned into a quantum point contact. The arrow highlights the anomalous peak in the differential resistance $R$  (equivalently to a differential conductance dip), from which $I_\textnormal{sw}$ is read off. To enhance the visibility of the above-gap anomaly, the color plot shows the numerical derivative of the differential resistance with respect to the DC current. The derivative is taken to remove the background resistance, which changes significantly as the QPC modes are sequentially opened. The orange triangles show the expected switching current according to the dissipation-based model as a function of mode number $N$, taking $P_\textnormal{c}\approx 0.5$ as a fitting parameter. The inset shows the dissipated power by the quantum point contact as a function of channel gate voltage.}
    \label{Fig3}
\end{figure}

In contrast to the voltage-biased measurements of the previous section, we probe this transition by measuring the differential resistance $R_\textnormal{diff}=dV/dI$ as a function of the current bias $I_\textnormal{DC}$. 
Figure~\ref{Fig3} shows the derivative $dR_\textnormal{diff}/dI_\textnormal{DC}$ as a function of $V_\textnormal{CG}$ and $I_\textnormal{DC}$, which enhances the visibility of the resistance peak associated with the superconducting-to-normal transition. We extract the switching current $I_\textnormal{sw}$ from the resulting feature in $dR_\textnormal{diff}/dI_\textnormal{DC}$.
Remarkably, the switching current evolves in a clear stepwise manner as individual transport modes are opened (see Fig.~\ref{Fig3}(a)). A similar stepwise increase of $I_\textnormal{sw}$ is observed in a nominally identical device (see Supplementary Material).

From the dissipation-based switching model, one expects $I_\textnormal{sw}=\sqrt{P_\textnormal{c} G}$ with $G=1/R$. The total conductance $G$ accounts for both the QPC channel and the parallel conduction paths identified earlier, $G=G_\textnormal{QPC}+G_{\parallel}$. Because in this regime the device conductance $G$ exhibits a stepwise evolution due to the QPC, the switching current should follow the quantized conductance $G_\textnormal{QPC}$.

For each mode number $N$, we extract the plateau value of $I_\textnormal{sw}$ and compare it to the model prediction, using $P_\textnormal{c}$ as a fit parameter. We obtain good agreement for $P_\textnormal{c}\approx~\SI{0.5}{nW}$, as shown in Fig.~\ref{Fig3}(b).
In Fig.~\ref{Fig3}(c), we further show the power dissipated at the switching point as a function of channel-gate voltage.
The dissipated power remains approximately constant at $P_c\approx\SI{0.5}{nW}$ throughout the sequential opening of transport modes, supporting a dissipation-driven transition.
This interpretation is further reinforced by benchmarking the $I_\textnormal{sw}\propto 1/\sqrt{R}$ scaling in the bulk regime and by the reproducibility of the quantized switching current in a second, nominally identical device (see Supplementary Information).


Having established that the low-bias conductance enhancement originates from Andreev reflection and that the above-gap anomaly is associated with the loss of superconductivity in the contact, we now quantify the associated excess current $I_\textnormal{exc}$, which provides a direct measure of the superconducting contribution to transport.

\begin{figure}[t]
    \centering
    \includegraphics[width=1\linewidth]{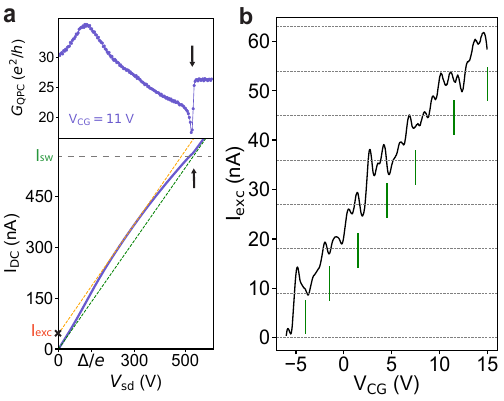}
    \caption{(a) Example of the extraction of the excess current for a fixed channel gate voltage  $V_\textnormal{CG}=\SI{11}{V}$. The top curves show the differential conductance, where an anomalous dip, indicated by the black arrow, marks the superconducting to normal transition.  The bottom plot shows the current-voltage curve, $I(V)$, obtained by integrating the QPC differential conductance shown in the top curve. The orange dashed line shows that as the bias voltage increases above the superconducting gap, the $I-V$ develops a current offset corresponding to $I_\textnormal{exc}$. The green dashed line indicates the normal quasiparticle branch, where no excess current is present. (b) Extracted excess current as a function of channel gate voltage. The vertical green lines indicate the position of the channel gate voltages corresponding to the center of the QPC conductance plateaus, identified from our previous measurements. }
    \label{fig4}
\end{figure}

We independently extract the excess current $I_\textnormal{exc}$ from the bias spectroscopy measurements, see the top curve in Fig.~\ref{Fig2}(c). This is achieved by integrating the measured differential conductance traces to reconstruct the current–voltage curves, where the excess current appears as an offset with respect to the normal-state current. In Fig.~\ref{fig4}(a), we show an example of such analysis, at a fixed channel gate voltage $V_\textnormal{CG}=\SI{11}{V}$. The top curve shows the differential conductance data, where the above-gap anomaly is indicated by the black arrow. In the bottom plot, we show the result of numerically integrating the top trace, with the integration constant fixed to $I=0$ at $V=0$. As predicted by BTK theory, the integrated $I-V$ curve for biases larger than the superconducting gap follows the expression in Eq.~\eqref{IV prediction}. The orange dashed line extrapolates the linear high-bias branch expected from Eq.~\eqref{IV prediction} down to zero bias. The current axis intercept of this line corresponds to the excess current. At larger biases, where the current reaches the switching current $I_\textnormal{sw}$, superconductivity in the contact and thereby also the excess current $I_\textnormal{exc}$ is lost. Beyond this point, the integrated curve follows $I=V/R$, indicated by the green dashed line, with $R$ being the normal state resistance. To avoid large extrapolation errors, we extract the excess current from the difference between the two dashed lines.

We repeat this procedure for different channel-gate voltages to determine how the excess current evolves as successive QPC modes are populated.
In Fig.~\ref{fig4}(b) we show the extracted $I_\textnormal{exc}$ as a function of $V_\textnormal{CG}$. The green vertical lines mark the channel gate voltages corresponding to the centers of the conductance plateaus identified in our previous measurements (see Fig.\ref{Fig2}(a,b), Fig.\ref{Fig3}(a)). $I_\textnormal{exc}$ increases as the constriction is opened and it is correlated with the sequential opening of transport modes, as emphasized by the horizontal dashed lines.

The excess current is predicted to scale proportionally to the superconducting gap and inversely proportional to the normal-state resistance $R$. In the clean limit $( \tau \approx~1)$, it is given by \cite{blonder1982transition, Review_PC_andreev}
\[ I_\textnormal{exc}=\frac{4}{3R} \frac{\Delta}{e}.\] For a QPC with $N$ transmitting modes, this expression becomes
\[ I_\textnormal{exc}=\frac{4}{3}\left(\frac{4e^2}{h}N\right) \frac{\Delta}{e},\]
which predicts increments of $\Delta I_{exc}\approx \SI{28}{nA}$ for a superconducting gap of \SI{120}{\micro V}. 
Experimentally, we observe increments of $\Delta I_\textnormal{exc}\approx 9\text{ nA}$ per mode.
This value is suppressed relative to the clean-limit prediction of $\SI{28}{nA}$ and the dirty-limit estimate of $\approx\SI{14}{nA}$.
Such a reduction is expected given our extracted interface transparency of $\tau\approx 0.88$, as well as the
elevated electronic temperatures at the high bias voltages $V_\textnormal{sd}>\SI{500}{\micro V}$ required for the extraction, which locally suppress the superconducting gap.

Despite this quantitative suppression, these results, together with the quantized switching current observed in Fig.~\ref{Fig3}, show that both equilibrium Andreev transport and the nonequilibrium loss of superconductivity inherit the mode structure of the ballistic channel. The observed steps occur at the same gate voltages as the conductance plateaus identified in Fig.~\ref{Fig1}, confirming that both $I_\text{sw}$ and $I_\text{exc}$
are governed by the sequential occupation of the fourfold-degenerate QPC modes.


In conclusion, we have realized a gate-defined bilayer graphene quantum point contact proximitized by a single superconducting aluminum electrode and investigated how discrete one-dimensional modes influence superconducting transport. We observe conductance plateaus enhanced by Andreev reflection, together with a  mode-dependent excess current and a switching current that evolves in discrete steps as successive QPC modes are populated. The switching current is consistent with a dissipation-driven transition occurring at an approximately constant critical power.

These results establish proximitized bilayer graphene QPCs as a versatile platform for exploring mode-resolved superconducting transport, with opportunities to investigate Andreev processes, dissipation mechanisms, and superconducting correlations in more complex ballistic nanostructures.

\section{Acknowledgements}
We thank Peter M\"{a}rki for support with the measurement electronics, and the staff at the ETH cleanroom FIRST for assistance with fabrication. We acknowledge discussions with Elias Portolés, Max Ruckriegel, Francesco Blanda, Fabrizio Volante, and Andrea Hofmann. This work was supported by the Quantum Center of ETH Zurich. C.G.A is grateful for financial support by the Heidi Ras Foundation.

%

\bibliography{Bibliography}

\begin{thebibliography}{54}%
\makeatletter
\providecommand \@ifxundefined [1]{%
 \@ifx{#1\undefined}
}%
\providecommand \@ifnum [1]{%
 \ifnum #1\expandafter \@firstoftwo
 \else \expandafter \@secondoftwo
 \fi
}%
\providecommand \@ifx [1]{%
 \ifx #1\expandafter \@firstoftwo
 \else \expandafter \@secondoftwo
 \fi
}%
\providecommand \natexlab [1]{#1}%
\providecommand \enquote  [1]{``#1''}%
\providecommand \bibnamefont  [1]{#1}%
\providecommand \bibfnamefont [1]{#1}%
\providecommand \citenamefont [1]{#1}%
\providecommand \href@noop [0]{\@secondoftwo}%
\providecommand \href [0]{\begingroup \@sanitize@url \@href}%
\providecommand \@href[1]{\@@startlink{#1}\@@href}%
\providecommand \@@href[1]{\endgroup#1\@@endlink}%
\providecommand \@sanitize@url [0]{\catcode `\\12\catcode `\$12\catcode `\&12\catcode `\#12\catcode `\^12\catcode `\_12\catcode `\%12\relax}%
\providecommand \@@startlink[1]{}%
\providecommand \@@endlink[0]{}%
\providecommand \url  [0]{\begingroup\@sanitize@url \@url }%
\providecommand \@url [1]{\endgroup\@href {#1}{\urlprefix }}%
\providecommand \urlprefix  [0]{URL }%
\providecommand \Eprint [0]{\href }%
\providecommand \doibase [0]{http://dx.doi.org/}%
\providecommand \selectlanguage [0]{\@gobble}%
\providecommand \bibinfo  [0]{\@secondoftwo}%
\providecommand \bibfield  [0]{\@secondoftwo}%
\providecommand \translation [1]{[#1]}%
\providecommand \BibitemOpen [0]{}%
\providecommand \bibitemStop [0]{}%
\providecommand \bibitemNoStop [0]{.\EOS\space}%
\providecommand \EOS [0]{\spacefactor3000\relax}%
\providecommand \BibitemShut  [1]{\csname bibitem#1\endcsname}%
\let\auto@bib@innerbib\@empty
\bibitem [{\citenamefont {Oostinga}\ \emph {et~al.}(2008)\citenamefont {Oostinga}, \citenamefont {Heersche}, \citenamefont {Liu}, \citenamefont {Morpurgo},\ and\ \citenamefont {Vandersypen}}]{oostinga2008gate}%
  \BibitemOpen
  \bibfield  {author} {\bibinfo {author} {\bibfnamefont {J.~B.}\ \bibnamefont {Oostinga}}, \bibinfo {author} {\bibfnamefont {H.~B.}\ \bibnamefont {Heersche}}, \bibinfo {author} {\bibfnamefont {X.}~\bibnamefont {Liu}}, \bibinfo {author} {\bibfnamefont {A.~F.}\ \bibnamefont {Morpurgo}}, \ and\ \bibinfo {author} {\bibfnamefont {L.~M.}\ \bibnamefont {Vandersypen}},\ }\href@noop {} {\bibfield  {journal} {\bibinfo  {journal} {Nat. Mater.}\ }\textbf {\bibinfo {volume} {7}},\ \bibinfo {pages} {151} (\bibinfo {year} {2008})}\BibitemShut {NoStop}%
\bibitem [{\citenamefont {Overweg}\ \emph {et~al.}(2018{\natexlab{a}})\citenamefont {Overweg}, \citenamefont {Eggimann}, \citenamefont {Chen}, \citenamefont {Slizovskiy}, \citenamefont {Eich}, \citenamefont {Pisoni}, \citenamefont {Lee}, \citenamefont {Rickhaus}, \citenamefont {Watanabe}, \citenamefont {Taniguchi} \emph {et~al.}}]{QPC_hiske}%
  \BibitemOpen
  \bibfield  {author} {\bibinfo {author} {\bibfnamefont {H.}~\bibnamefont {Overweg}}, \bibinfo {author} {\bibfnamefont {H.}~\bibnamefont {Eggimann}}, \bibinfo {author} {\bibfnamefont {X.}~\bibnamefont {Chen}}, \bibinfo {author} {\bibfnamefont {S.}~\bibnamefont {Slizovskiy}}, \bibinfo {author} {\bibfnamefont {M.}~\bibnamefont {Eich}}, \bibinfo {author} {\bibfnamefont {R.}~\bibnamefont {Pisoni}}, \bibinfo {author} {\bibfnamefont {Y.}~\bibnamefont {Lee}}, \bibinfo {author} {\bibfnamefont {P.}~\bibnamefont {Rickhaus}}, \bibinfo {author} {\bibfnamefont {K.}~\bibnamefont {Watanabe}}, \bibinfo {author} {\bibfnamefont {T.}~\bibnamefont {Taniguchi}},  \emph {et~al.},\ }\href@noop {} {\bibfield  {journal} {\bibinfo  {journal} {Nano Lett.}\ }\textbf {\bibinfo {volume} {18}},\ \bibinfo {pages} {553} (\bibinfo {year} {2018}{\natexlab{a}})}\BibitemShut {NoStop}%
\bibitem [{\citenamefont {Banszerus}\ \emph {et~al.}(2020)\citenamefont {Banszerus}, \citenamefont {Frohn}, \citenamefont {Fabian}, \citenamefont {Somanchi}, \citenamefont {Epping}, \citenamefont {M\"uller}, \citenamefont {Neumaier}, \citenamefont {Watanabe}, \citenamefont {Taniguchi}, \citenamefont {Libisch}, \citenamefont {Beschoten}, \citenamefont {Hassler},\ and\ \citenamefont {Stampfer}}]{PhysRevLett.124.177701}%
  \BibitemOpen
  \bibfield  {author} {\bibinfo {author} {\bibfnamefont {L.}~\bibnamefont {Banszerus}}, \bibinfo {author} {\bibfnamefont {B.}~\bibnamefont {Frohn}}, \bibinfo {author} {\bibfnamefont {T.}~\bibnamefont {Fabian}}, \bibinfo {author} {\bibfnamefont {S.}~\bibnamefont {Somanchi}}, \bibinfo {author} {\bibfnamefont {A.}~\bibnamefont {Epping}}, \bibinfo {author} {\bibfnamefont {M.}~\bibnamefont {M\"uller}}, \bibinfo {author} {\bibfnamefont {D.}~\bibnamefont {Neumaier}}, \bibinfo {author} {\bibfnamefont {K.}~\bibnamefont {Watanabe}}, \bibinfo {author} {\bibfnamefont {T.}~\bibnamefont {Taniguchi}}, \bibinfo {author} {\bibfnamefont {F.}~\bibnamefont {Libisch}}, \bibinfo {author} {\bibfnamefont {B.}~\bibnamefont {Beschoten}}, \bibinfo {author} {\bibfnamefont {F.}~\bibnamefont {Hassler}}, \ and\ \bibinfo {author} {\bibfnamefont {C.}~\bibnamefont {Stampfer}},\ }\href {\doibase 10.1103/PhysRevLett.124.177701} {\bibfield  {journal} {\bibinfo  {journal} {Phys. Rev. Lett.}\ }\textbf {\bibinfo {volume} {124}},\ \bibinfo {pages}
  {177701} (\bibinfo {year} {2020})}\BibitemShut {NoStop}%
\bibitem [{\citenamefont {Lee}\ \emph {et~al.}(2020)\citenamefont {Lee}, \citenamefont {Knothe}, \citenamefont {Overweg}, \citenamefont {Eich}, \citenamefont {Gold}, \citenamefont {Kurzmann}, \citenamefont {Klasovika}, \citenamefont {Taniguchi}, \citenamefont {Wantanabe}, \citenamefont {Fal'ko} \emph {et~al.}}]{lee2020tunable}%
  \BibitemOpen
  \bibfield  {author} {\bibinfo {author} {\bibfnamefont {Y.}~\bibnamefont {Lee}}, \bibinfo {author} {\bibfnamefont {A.}~\bibnamefont {Knothe}}, \bibinfo {author} {\bibfnamefont {H.}~\bibnamefont {Overweg}}, \bibinfo {author} {\bibfnamefont {M.}~\bibnamefont {Eich}}, \bibinfo {author} {\bibfnamefont {C.}~\bibnamefont {Gold}}, \bibinfo {author} {\bibfnamefont {A.}~\bibnamefont {Kurzmann}}, \bibinfo {author} {\bibfnamefont {V.}~\bibnamefont {Klasovika}}, \bibinfo {author} {\bibfnamefont {T.}~\bibnamefont {Taniguchi}}, \bibinfo {author} {\bibfnamefont {K.}~\bibnamefont {Wantanabe}}, \bibinfo {author} {\bibfnamefont {V.}~\bibnamefont {Fal'ko}},  \emph {et~al.},\ }\href@noop {} {\bibfield  {journal} {\bibinfo  {journal} {Phys. Rev. Lett.}\ }\textbf {\bibinfo {volume} {124}},\ \bibinfo {pages} {126802} (\bibinfo {year} {2020})}\BibitemShut {NoStop}%
\bibitem [{\citenamefont {Eich}\ \emph {et~al.}(2018{\natexlab{a}})\citenamefont {Eich}, \citenamefont {Herman}, \citenamefont {Pisoni}, \citenamefont {Overweg}, \citenamefont {Kurzmann}, \citenamefont {Lee}, \citenamefont {Rickhaus}, \citenamefont {Watanabe}, \citenamefont {Taniguchi}, \citenamefont {Sigrist} \emph {et~al.}}]{eich2018spin}%
  \BibitemOpen
  \bibfield  {author} {\bibinfo {author} {\bibfnamefont {M.}~\bibnamefont {Eich}}, \bibinfo {author} {\bibfnamefont {F.}~\bibnamefont {Herman}}, \bibinfo {author} {\bibfnamefont {R.}~\bibnamefont {Pisoni}}, \bibinfo {author} {\bibfnamefont {H.}~\bibnamefont {Overweg}}, \bibinfo {author} {\bibfnamefont {A.}~\bibnamefont {Kurzmann}}, \bibinfo {author} {\bibfnamefont {Y.}~\bibnamefont {Lee}}, \bibinfo {author} {\bibfnamefont {P.}~\bibnamefont {Rickhaus}}, \bibinfo {author} {\bibfnamefont {K.}~\bibnamefont {Watanabe}}, \bibinfo {author} {\bibfnamefont {T.}~\bibnamefont {Taniguchi}}, \bibinfo {author} {\bibfnamefont {M.}~\bibnamefont {Sigrist}},  \emph {et~al.},\ }\href@noop {} {\bibfield  {journal} {\bibinfo  {journal} {Phys. Rev. X}\ }\textbf {\bibinfo {volume} {8}},\ \bibinfo {pages} {031023} (\bibinfo {year} {2018}{\natexlab{a}})}\BibitemShut {NoStop}%
\bibitem [{\citenamefont {Banszerus}\ \emph {et~al.}(2018)\citenamefont {Banszerus}, \citenamefont {Frohn}, \citenamefont {Epping}, \citenamefont {Neumaier}, \citenamefont {Watanabe}, \citenamefont {Taniguchi},\ and\ \citenamefont {Stampfer}}]{Banszerus18}%
  \BibitemOpen
  \bibfield  {author} {\bibinfo {author} {\bibfnamefont {L.}~\bibnamefont {Banszerus}}, \bibinfo {author} {\bibfnamefont {B.}~\bibnamefont {Frohn}}, \bibinfo {author} {\bibfnamefont {A.}~\bibnamefont {Epping}}, \bibinfo {author} {\bibfnamefont {D.}~\bibnamefont {Neumaier}}, \bibinfo {author} {\bibfnamefont {K.}~\bibnamefont {Watanabe}}, \bibinfo {author} {\bibfnamefont {T.}~\bibnamefont {Taniguchi}}, \ and\ \bibinfo {author} {\bibfnamefont {C.}~\bibnamefont {Stampfer}},\ }\href@noop {} {\bibfield  {journal} {\bibinfo  {journal} {Nano Lett.}\ }\textbf {\bibinfo {volume} {18}},\ \bibinfo {pages} {4785} (\bibinfo {year} {2018})}\BibitemShut {NoStop}%
\bibitem [{\citenamefont {Eich}\ \emph {et~al.}(2018{\natexlab{b}})\citenamefont {Eich}, \citenamefont {Pisoni}, \citenamefont {Pally}, \citenamefont {Overweg}, \citenamefont {Kurzmann}, \citenamefont {Lee}, \citenamefont {Rickhaus}, \citenamefont {Watanabe}, \citenamefont {Taniguchi}, \citenamefont {Ensslin},\ and\ \citenamefont {Ihn}}]{Eich18}%
  \BibitemOpen
  \bibfield  {author} {\bibinfo {author} {\bibfnamefont {M.}~\bibnamefont {Eich}}, \bibinfo {author} {\bibfnamefont {R.}~\bibnamefont {Pisoni}}, \bibinfo {author} {\bibfnamefont {A.}~\bibnamefont {Pally}}, \bibinfo {author} {\bibfnamefont {H.}~\bibnamefont {Overweg}}, \bibinfo {author} {\bibfnamefont {A.}~\bibnamefont {Kurzmann}}, \bibinfo {author} {\bibfnamefont {Y.}~\bibnamefont {Lee}}, \bibinfo {author} {\bibfnamefont {P.}~\bibnamefont {Rickhaus}}, \bibinfo {author} {\bibfnamefont {K.}~\bibnamefont {Watanabe}}, \bibinfo {author} {\bibfnamefont {T.}~\bibnamefont {Taniguchi}}, \bibinfo {author} {\bibfnamefont {K.}~\bibnamefont {Ensslin}}, \ and\ \bibinfo {author} {\bibfnamefont {T.}~\bibnamefont {Ihn}},\ }\href@noop {} {\bibfield  {journal} {\bibinfo  {journal} {Nano Lett.}\ }\textbf {\bibinfo {volume} {18}},\ \bibinfo {pages} {5042} (\bibinfo {year} {2018}{\natexlab{b}})}\BibitemShut {NoStop}%
\bibitem [{\citenamefont {Kurzmann}\ \emph {et~al.}(2019)\citenamefont {Kurzmann}, \citenamefont {Eich}, \citenamefont {Overweg}, \citenamefont {Mangold}, \citenamefont {Herman}, \citenamefont {Rickhaus}, \citenamefont {Pisoni}, \citenamefont {Lee}, \citenamefont {Garreis}, \citenamefont {Tong}, \citenamefont {Watanabe}, \citenamefont {Taniguchi}, \citenamefont {Ensslin},\ and\ \citenamefont {Ihn}}]{PhysRevLett.123.026803}%
  \BibitemOpen
  \bibfield  {author} {\bibinfo {author} {\bibfnamefont {A.}~\bibnamefont {Kurzmann}}, \bibinfo {author} {\bibfnamefont {M.}~\bibnamefont {Eich}}, \bibinfo {author} {\bibfnamefont {H.}~\bibnamefont {Overweg}}, \bibinfo {author} {\bibfnamefont {M.}~\bibnamefont {Mangold}}, \bibinfo {author} {\bibfnamefont {F.}~\bibnamefont {Herman}}, \bibinfo {author} {\bibfnamefont {P.}~\bibnamefont {Rickhaus}}, \bibinfo {author} {\bibfnamefont {R.}~\bibnamefont {Pisoni}}, \bibinfo {author} {\bibfnamefont {Y.}~\bibnamefont {Lee}}, \bibinfo {author} {\bibfnamefont {R.}~\bibnamefont {Garreis}}, \bibinfo {author} {\bibfnamefont {C.}~\bibnamefont {Tong}}, \bibinfo {author} {\bibfnamefont {K.}~\bibnamefont {Watanabe}}, \bibinfo {author} {\bibfnamefont {T.}~\bibnamefont {Taniguchi}}, \bibinfo {author} {\bibfnamefont {K.}~\bibnamefont {Ensslin}}, \ and\ \bibinfo {author} {\bibfnamefont {T.}~\bibnamefont {Ihn}},\ }\href {\doibase 10.1103/PhysRevLett.123.026803} {\bibfield  {journal} {\bibinfo  {journal} {Phys. Rev. Lett.}\ }\textbf
  {\bibinfo {volume} {123}},\ \bibinfo {pages} {026803} (\bibinfo {year} {2019})}\BibitemShut {NoStop}%
\bibitem [{\citenamefont {Li}\ \emph {et~al.}(2018)\citenamefont {Li}, \citenamefont {Zhang}, \citenamefont {Yin}, \citenamefont {Zhang}, \citenamefont {Watanabe}, \citenamefont {Taniguchi}, \citenamefont {Liu},\ and\ \citenamefont {Zhu}}]{valley_valve}%
  \BibitemOpen
  \bibfield  {author} {\bibinfo {author} {\bibfnamefont {J.}~\bibnamefont {Li}}, \bibinfo {author} {\bibfnamefont {R.-X.}\ \bibnamefont {Zhang}}, \bibinfo {author} {\bibfnamefont {Z.}~\bibnamefont {Yin}}, \bibinfo {author} {\bibfnamefont {J.}~\bibnamefont {Zhang}}, \bibinfo {author} {\bibfnamefont {K.}~\bibnamefont {Watanabe}}, \bibinfo {author} {\bibfnamefont {T.}~\bibnamefont {Taniguchi}}, \bibinfo {author} {\bibfnamefont {C.}~\bibnamefont {Liu}}, \ and\ \bibinfo {author} {\bibfnamefont {J.}~\bibnamefont {Zhu}},\ }\href@noop {} {\bibfield  {journal} {\bibinfo  {journal} {Science}\ }\textbf {\bibinfo {volume} {362}},\ \bibinfo {pages} {1149} (\bibinfo {year} {2018})}\BibitemShut {NoStop}%
\bibitem [{\citenamefont {Chen}\ \emph {et~al.}(2020)\citenamefont {Chen}, \citenamefont {Zhou}, \citenamefont {Liu}, \citenamefont {Qiao}, \citenamefont {Oezyilmaz},\ and\ \citenamefont {Martin}}]{valley_polarizer}%
  \BibitemOpen
  \bibfield  {author} {\bibinfo {author} {\bibfnamefont {H.}~\bibnamefont {Chen}}, \bibinfo {author} {\bibfnamefont {P.}~\bibnamefont {Zhou}}, \bibinfo {author} {\bibfnamefont {J.}~\bibnamefont {Liu}}, \bibinfo {author} {\bibfnamefont {J.}~\bibnamefont {Qiao}}, \bibinfo {author} {\bibfnamefont {B.}~\bibnamefont {Oezyilmaz}}, \ and\ \bibinfo {author} {\bibfnamefont {J.}~\bibnamefont {Martin}},\ }\href@noop {} {\bibfield  {journal} {\bibinfo  {journal} {Nat. Commun.}\ }\textbf {\bibinfo {volume} {11}},\ \bibinfo {pages} {1202} (\bibinfo {year} {2020})}\BibitemShut {NoStop}%
\bibitem [{\citenamefont {Varlet}\ \emph {et~al.}(2014)\citenamefont {Varlet}, \citenamefont {Liu}, \citenamefont {Krueckl}, \citenamefont {Bischoff}, \citenamefont {Simonet}, \citenamefont {Watanabe}, \citenamefont {Taniguchi}, \citenamefont {Richter}, \citenamefont {Ensslin},\ and\ \citenamefont {Ihn}}]{Fabry_perot}%
  \BibitemOpen
  \bibfield  {author} {\bibinfo {author} {\bibfnamefont {A.}~\bibnamefont {Varlet}}, \bibinfo {author} {\bibfnamefont {M.-H.}\ \bibnamefont {Liu}}, \bibinfo {author} {\bibfnamefont {V.}~\bibnamefont {Krueckl}}, \bibinfo {author} {\bibfnamefont {D.}~\bibnamefont {Bischoff}}, \bibinfo {author} {\bibfnamefont {P.}~\bibnamefont {Simonet}}, \bibinfo {author} {\bibfnamefont {K.}~\bibnamefont {Watanabe}}, \bibinfo {author} {\bibfnamefont {T.}~\bibnamefont {Taniguchi}}, \bibinfo {author} {\bibfnamefont {K.}~\bibnamefont {Richter}}, \bibinfo {author} {\bibfnamefont {K.}~\bibnamefont {Ensslin}}, \ and\ \bibinfo {author} {\bibfnamefont {T.}~\bibnamefont {Ihn}},\ }\href {\doibase 10.1103/PhysRevLett.113.116601} {\bibfield  {journal} {\bibinfo  {journal} {Phys. Rev. Lett.}\ }\textbf {\bibinfo {volume} {113}},\ \bibinfo {pages} {116601} (\bibinfo {year} {2014})}\BibitemShut {NoStop}%
\bibitem [{\citenamefont {Fu}\ \emph {et~al.}(2023)\citenamefont {Fu}, \citenamefont {Huang}, \citenamefont {Watanabe}, \citenamefont {Taniguchi}, \citenamefont {Kayyalha},\ and\ \citenamefont {Zhu}}]{AB_oscillations}%
  \BibitemOpen
  \bibfield  {author} {\bibinfo {author} {\bibfnamefont {H.}~\bibnamefont {Fu}}, \bibinfo {author} {\bibfnamefont {K.}~\bibnamefont {Huang}}, \bibinfo {author} {\bibfnamefont {K.}~\bibnamefont {Watanabe}}, \bibinfo {author} {\bibfnamefont {T.}~\bibnamefont {Taniguchi}}, \bibinfo {author} {\bibfnamefont {M.}~\bibnamefont {Kayyalha}}, \ and\ \bibinfo {author} {\bibfnamefont {J.}~\bibnamefont {Zhu}},\ }\href {\doibase 10.1021/acs.nanolett.2c05004} {\bibfield  {journal} {\bibinfo  {journal} {Nano Letters}\ }\textbf {\bibinfo {volume} {23}},\ \bibinfo {pages} {718} (\bibinfo {year} {2023})}\BibitemShut {NoStop}%
\bibitem [{\citenamefont {D{\'e}prez}\ \emph {et~al.}(2021)\citenamefont {D{\'e}prez}, \citenamefont {Veyrat}, \citenamefont {Vignaud}, \citenamefont {Nayak}, \citenamefont {Watanabe}, \citenamefont {Taniguchi}, \citenamefont {Gay}, \citenamefont {Sellier},\ and\ \citenamefont {Sac{\'e}p{\'e}}}]{Deprez2021}%
  \BibitemOpen
  \bibfield  {author} {\bibinfo {author} {\bibfnamefont {C.}~\bibnamefont {D{\'e}prez}}, \bibinfo {author} {\bibfnamefont {L.}~\bibnamefont {Veyrat}}, \bibinfo {author} {\bibfnamefont {H.}~\bibnamefont {Vignaud}}, \bibinfo {author} {\bibfnamefont {G.}~\bibnamefont {Nayak}}, \bibinfo {author} {\bibfnamefont {K.}~\bibnamefont {Watanabe}}, \bibinfo {author} {\bibfnamefont {T.}~\bibnamefont {Taniguchi}}, \bibinfo {author} {\bibfnamefont {F.}~\bibnamefont {Gay}}, \bibinfo {author} {\bibfnamefont {H.}~\bibnamefont {Sellier}}, \ and\ \bibinfo {author} {\bibfnamefont {B.}~\bibnamefont {Sac{\'e}p{\'e}}},\ }\href {\doibase 10.1038/s41565--021--00847--x} {\bibfield  {journal} {\bibinfo  {journal} {Nature Nanotechnology}\ }\textbf {\bibinfo {volume} {16}},\ \bibinfo {pages} {555} (\bibinfo {year} {2021})}\BibitemShut {NoStop}%
\bibitem [{\citenamefont {Beenakker}(2006)}]{PhysRevLett.97.067007}%
  \BibitemOpen
  \bibfield  {author} {\bibinfo {author} {\bibfnamefont {C.~W.~J.}\ \bibnamefont {Beenakker}},\ }\href {\doibase 10.1103/PhysRevLett.97.067007} {\bibfield  {journal} {\bibinfo  {journal} {Phys. Rev. Lett.}\ }\textbf {\bibinfo {volume} {97}},\ \bibinfo {pages} {067007} (\bibinfo {year} {2006})}\BibitemShut {NoStop}%
\bibitem [{\citenamefont {Heersche}\ \emph {et~al.}(2007)\citenamefont {Heersche}, \citenamefont {Jarillo-Herrero}, \citenamefont {Oostinga}, \citenamefont {Vandersypen},\ and\ \citenamefont {Morpurgo}}]{bipolar_supercurrent_Pablo}%
  \BibitemOpen
  \bibfield  {author} {\bibinfo {author} {\bibfnamefont {H.~B.}\ \bibnamefont {Heersche}}, \bibinfo {author} {\bibfnamefont {P.}~\bibnamefont {Jarillo-Herrero}}, \bibinfo {author} {\bibfnamefont {J.~B.}\ \bibnamefont {Oostinga}}, \bibinfo {author} {\bibfnamefont {L.~M.}\ \bibnamefont {Vandersypen}}, \ and\ \bibinfo {author} {\bibfnamefont {A.~F.}\ \bibnamefont {Morpurgo}},\ }\href@noop {} {\bibfield  {journal} {\bibinfo  {journal} {Nature}\ }\textbf {\bibinfo {volume} {446}},\ \bibinfo {pages} {56} (\bibinfo {year} {2007})}\BibitemShut {NoStop}%
\bibitem [{\citenamefont {Sato}\ \emph {et~al.}(2008)\citenamefont {Sato}, \citenamefont {Moriki}, \citenamefont {Tanakaa}, \citenamefont {Kanda}, \citenamefont {Goto}, \citenamefont {Miyazaki}, \citenamefont {Odaka}, \citenamefont {Ootuka}, \citenamefont {Tsukagoshi},\ and\ \citenamefont {Aoyagi}}]{Sato07}%
  \BibitemOpen
  \bibfield  {author} {\bibinfo {author} {\bibfnamefont {T.}~\bibnamefont {Sato}}, \bibinfo {author} {\bibfnamefont {T.}~\bibnamefont {Moriki}}, \bibinfo {author} {\bibfnamefont {S.}~\bibnamefont {Tanakaa}}, \bibinfo {author} {\bibfnamefont {A.}~\bibnamefont {Kanda}}, \bibinfo {author} {\bibfnamefont {H.}~\bibnamefont {Goto}}, \bibinfo {author} {\bibfnamefont {H.}~\bibnamefont {Miyazaki}}, \bibinfo {author} {\bibfnamefont {S.}~\bibnamefont {Odaka}}, \bibinfo {author} {\bibfnamefont {Y.}~\bibnamefont {Ootuka}}, \bibinfo {author} {\bibfnamefont {K.}~\bibnamefont {Tsukagoshi}}, \ and\ \bibinfo {author} {\bibfnamefont {Y.}~\bibnamefont {Aoyagi}},\ }\href@noop {} {\bibfield  {journal} {\bibinfo  {journal} {Physica E}\ }\textbf {\bibinfo {volume} {40}},\ \bibinfo {pages} {1495} (\bibinfo {year} {2008})}\BibitemShut {NoStop}%
\bibitem [{\citenamefont {Du}\ \emph {et~al.}(2008{\natexlab{a}})\citenamefont {Du}, \citenamefont {Skachko},\ and\ \citenamefont {Andrei}}]{PhysRevB.77.184507}%
  \BibitemOpen
  \bibfield  {author} {\bibinfo {author} {\bibfnamefont {X.}~\bibnamefont {Du}}, \bibinfo {author} {\bibfnamefont {I.}~\bibnamefont {Skachko}}, \ and\ \bibinfo {author} {\bibfnamefont {E.~Y.}\ \bibnamefont {Andrei}},\ }\href {\doibase 10.1103/PhysRevB.77.184507} {\bibfield  {journal} {\bibinfo  {journal} {Phys. Rev. B}\ }\textbf {\bibinfo {volume} {77}},\ \bibinfo {pages} {184507} (\bibinfo {year} {2008}{\natexlab{a}})}\BibitemShut {NoStop}%
\bibitem [{\citenamefont {Ojeda-Aristizabal}\ \emph {et~al.}(2009)\citenamefont {Ojeda-Aristizabal}, \citenamefont {Ferrier}, \citenamefont {Gu\'eron},\ and\ \citenamefont {Bouchiat}}]{PhysRevB.79.165436}%
  \BibitemOpen
  \bibfield  {author} {\bibinfo {author} {\bibfnamefont {C.}~\bibnamefont {Ojeda-Aristizabal}}, \bibinfo {author} {\bibfnamefont {M.}~\bibnamefont {Ferrier}}, \bibinfo {author} {\bibfnamefont {S.}~\bibnamefont {Gu\'eron}}, \ and\ \bibinfo {author} {\bibfnamefont {H.}~\bibnamefont {Bouchiat}},\ }\href {\doibase 10.1103/PhysRevB.79.165436} {\bibfield  {journal} {\bibinfo  {journal} {Phys. Rev. B}\ }\textbf {\bibinfo {volume} {79}},\ \bibinfo {pages} {165436} (\bibinfo {year} {2009})}\BibitemShut {NoStop}%
\bibitem [{\citenamefont {Choi}\ \emph {et~al.}(2013)\citenamefont {Choi}, \citenamefont {Lee}, \citenamefont {Park}, \citenamefont {Jeong}, \citenamefont {Lee}, \citenamefont {Sim}, \citenamefont {Doh},\ and\ \citenamefont {Lee}}]{Choi13}%
  \BibitemOpen
  \bibfield  {author} {\bibinfo {author} {\bibfnamefont {J.-H.}\ \bibnamefont {Choi}}, \bibinfo {author} {\bibfnamefont {G.-H.}\ \bibnamefont {Lee}}, \bibinfo {author} {\bibfnamefont {S.}~\bibnamefont {Park}}, \bibinfo {author} {\bibfnamefont {D.}~\bibnamefont {Jeong}}, \bibinfo {author} {\bibfnamefont {J.-O.}\ \bibnamefont {Lee}}, \bibinfo {author} {\bibfnamefont {H.-S.}\ \bibnamefont {Sim}}, \bibinfo {author} {\bibfnamefont {Y.-J.}\ \bibnamefont {Doh}}, \ and\ \bibinfo {author} {\bibfnamefont {H.-J.}\ \bibnamefont {Lee}},\ }\href@noop {} {\bibfield  {journal} {\bibinfo  {journal} {Nat. Commun.}\ }\textbf {\bibinfo {volume} {4}},\ \bibinfo {pages} {2525} (\bibinfo {year} {2013})}\BibitemShut {NoStop}%
\bibitem [{\citenamefont {Lee}\ and\ \citenamefont {Lee}(2018)}]{lee2018proximity}%
  \BibitemOpen
  \bibfield  {author} {\bibinfo {author} {\bibfnamefont {G.-H.}\ \bibnamefont {Lee}}\ and\ \bibinfo {author} {\bibfnamefont {H.-J.}\ \bibnamefont {Lee}},\ }\href@noop {} {\bibfield  {journal} {\bibinfo  {journal} {Rep. Prog. Phys.}\ }\textbf {\bibinfo {volume} {81}},\ \bibinfo {pages} {056502} (\bibinfo {year} {2018})}\BibitemShut {NoStop}%
\bibitem [{\citenamefont {Calado}\ \emph {et~al.}(2015)\citenamefont {Calado}, \citenamefont {Goswami}, \citenamefont {Nanda}, \citenamefont {Diez}, \citenamefont {Akhmerov}, \citenamefont {Watanabe}, \citenamefont {Taniguchi}, \citenamefont {Klapwijk},\ and\ \citenamefont {Vandersypen}}]{Calado2015}%
  \BibitemOpen
  \bibfield  {author} {\bibinfo {author} {\bibfnamefont {V.~E.}\ \bibnamefont {Calado}}, \bibinfo {author} {\bibfnamefont {S.}~\bibnamefont {Goswami}}, \bibinfo {author} {\bibfnamefont {G.}~\bibnamefont {Nanda}}, \bibinfo {author} {\bibfnamefont {M.}~\bibnamefont {Diez}}, \bibinfo {author} {\bibfnamefont {A.~R.}\ \bibnamefont {Akhmerov}}, \bibinfo {author} {\bibfnamefont {K.}~\bibnamefont {Watanabe}}, \bibinfo {author} {\bibfnamefont {T.}~\bibnamefont {Taniguchi}}, \bibinfo {author} {\bibfnamefont {T.~M.}\ \bibnamefont {Klapwijk}}, \ and\ \bibinfo {author} {\bibfnamefont {L.~M.~K.}\ \bibnamefont {Vandersypen}},\ }\href {\doibase 10.1038/nnano.2015.156} {\bibfield  {journal} {\bibinfo  {journal} {Nat. Nanotechnol.}\ }\textbf {\bibinfo {volume} {10}},\ \bibinfo {pages} {761} (\bibinfo {year} {2015})}\BibitemShut {NoStop}%
\bibitem [{\citenamefont {Ben~Shalom}\ \emph {et~al.}(2016)\citenamefont {Ben~Shalom}, \citenamefont {Zhu}, \citenamefont {Fal'ko}, \citenamefont {Mishchenko}, \citenamefont {Kretinin}, \citenamefont {Novoselov}, \citenamefont {Woods}, \citenamefont {Watanabe}, \citenamefont {Taniguchi}, \citenamefont {Geim},\ and\ \citenamefont {Prance}}]{Quantum_oscillations_JJ}%
  \BibitemOpen
  \bibfield  {author} {\bibinfo {author} {\bibfnamefont {M.}~\bibnamefont {Ben~Shalom}}, \bibinfo {author} {\bibfnamefont {M.~J.}\ \bibnamefont {Zhu}}, \bibinfo {author} {\bibfnamefont {V.~I.}\ \bibnamefont {Fal'ko}}, \bibinfo {author} {\bibfnamefont {A.}~\bibnamefont {Mishchenko}}, \bibinfo {author} {\bibfnamefont {A.~V.}\ \bibnamefont {Kretinin}}, \bibinfo {author} {\bibfnamefont {K.~S.}\ \bibnamefont {Novoselov}}, \bibinfo {author} {\bibfnamefont {C.~R.}\ \bibnamefont {Woods}}, \bibinfo {author} {\bibfnamefont {K.}~\bibnamefont {Watanabe}}, \bibinfo {author} {\bibfnamefont {T.}~\bibnamefont {Taniguchi}}, \bibinfo {author} {\bibfnamefont {A.~K.}\ \bibnamefont {Geim}}, \ and\ \bibinfo {author} {\bibfnamefont {J.~R.}\ \bibnamefont {Prance}},\ }\href {\doibase 10.1038/nphys3592} {\bibfield  {journal} {\bibinfo  {journal} {Nat. Phys.}\ }\textbf {\bibinfo {volume} {12}},\ \bibinfo {pages} {318} (\bibinfo {year} {2016})}\BibitemShut {NoStop}%
\bibitem [{\citenamefont {Amet}\ \emph {et~al.}(2016)\citenamefont {Amet}, \citenamefont {Ke}, \citenamefont {Borzenets}, \citenamefont {Wang}, \citenamefont {Watanabe}, \citenamefont {Taniguchi}, \citenamefont {Deacon}, \citenamefont {Yamamoto}, \citenamefont {Bomze}, \citenamefont {Tarucha},\ and\ \citenamefont {Finkelstein}}]{S_and_quantum_hall}%
  \BibitemOpen
  \bibfield  {author} {\bibinfo {author} {\bibfnamefont {F.}~\bibnamefont {Amet}}, \bibinfo {author} {\bibfnamefont {C.~T.}\ \bibnamefont {Ke}}, \bibinfo {author} {\bibfnamefont {I.~V.}\ \bibnamefont {Borzenets}}, \bibinfo {author} {\bibfnamefont {J.}~\bibnamefont {Wang}}, \bibinfo {author} {\bibfnamefont {K.}~\bibnamefont {Watanabe}}, \bibinfo {author} {\bibfnamefont {T.}~\bibnamefont {Taniguchi}}, \bibinfo {author} {\bibfnamefont {R.~S.}\ \bibnamefont {Deacon}}, \bibinfo {author} {\bibfnamefont {M.}~\bibnamefont {Yamamoto}}, \bibinfo {author} {\bibfnamefont {Y.}~\bibnamefont {Bomze}}, \bibinfo {author} {\bibfnamefont {S.}~\bibnamefont {Tarucha}}, \ and\ \bibinfo {author} {\bibfnamefont {G.}~\bibnamefont {Finkelstein}},\ }\href {\doibase 10.1126/science.aad6203} {\bibfield  {journal} {\bibinfo  {journal} {Science}\ }\textbf {\bibinfo {volume} {352}},\ \bibinfo {pages} {966} (\bibinfo {year} {2016})},\ \Eprint {http://arxiv.org/abs/https://www.science.org/doi/pdf/10.1126/science.aad6203}
  {https://www.science.org/doi/pdf/10.1126/science.aad6203} \BibitemShut {NoStop}%
\bibitem [{\citenamefont {Kraft}\ \emph {et~al.}(2018)\citenamefont {Kraft}, \citenamefont {Mohrmann}, \citenamefont {Du}, \citenamefont {Selvasundaram}, \citenamefont {Irfan}, \citenamefont {Kanilmaz}, \citenamefont {Wu}, \citenamefont {Beckmann}, \citenamefont {von L{\"o}hneysen}, \citenamefont {Krupke}, \citenamefont {Akhmerov}, \citenamefont {Gornyi},\ and\ \citenamefont {Danneau}}]{Tayloring_supercurrents}%
  \BibitemOpen
  \bibfield  {author} {\bibinfo {author} {\bibfnamefont {R.}~\bibnamefont {Kraft}}, \bibinfo {author} {\bibfnamefont {J.}~\bibnamefont {Mohrmann}}, \bibinfo {author} {\bibfnamefont {R.}~\bibnamefont {Du}}, \bibinfo {author} {\bibfnamefont {P.~B.}\ \bibnamefont {Selvasundaram}}, \bibinfo {author} {\bibfnamefont {M.}~\bibnamefont {Irfan}}, \bibinfo {author} {\bibfnamefont {U.~N.}\ \bibnamefont {Kanilmaz}}, \bibinfo {author} {\bibfnamefont {F.}~\bibnamefont {Wu}}, \bibinfo {author} {\bibfnamefont {D.}~\bibnamefont {Beckmann}}, \bibinfo {author} {\bibfnamefont {H.}~\bibnamefont {von L{\"o}hneysen}}, \bibinfo {author} {\bibfnamefont {R.}~\bibnamefont {Krupke}}, \bibinfo {author} {\bibfnamefont {A.}~\bibnamefont {Akhmerov}}, \bibinfo {author} {\bibfnamefont {I.}~\bibnamefont {Gornyi}}, \ and\ \bibinfo {author} {\bibfnamefont {R.}~\bibnamefont {Danneau}},\ }\href {\doibase 10.1038/s41467--018--04153--4} {\bibfield  {journal} {\bibinfo  {journal} {Nat. Commun.}\ }\textbf {\bibinfo {volume} {9}},\ \bibinfo {pages}
  {1722} (\bibinfo {year} {2018})}\BibitemShut {NoStop}%
\bibitem [{\citenamefont {Park}\ \emph {et~al.}(2024)\citenamefont {Park}, \citenamefont {Lee}, \citenamefont {Park}, \citenamefont {Watanabe}, \citenamefont {Taniguchi}, \citenamefont {Cho},\ and\ \citenamefont {Lee}}]{ABS_resonator}%
  \BibitemOpen
  \bibfield  {author} {\bibinfo {author} {\bibfnamefont {G.-H.}\ \bibnamefont {Park}}, \bibinfo {author} {\bibfnamefont {W.}~\bibnamefont {Lee}}, \bibinfo {author} {\bibfnamefont {S.}~\bibnamefont {Park}}, \bibinfo {author} {\bibfnamefont {K.}~\bibnamefont {Watanabe}}, \bibinfo {author} {\bibfnamefont {T.}~\bibnamefont {Taniguchi}}, \bibinfo {author} {\bibfnamefont {G.~Y.}\ \bibnamefont {Cho}}, \ and\ \bibinfo {author} {\bibfnamefont {G.-H.}\ \bibnamefont {Lee}},\ }\href {\doibase 10.1103/PhysRevLett.132.226301} {\bibfield  {journal} {\bibinfo  {journal} {Phys. Rev. Lett.}\ }\textbf {\bibinfo {volume} {132}},\ \bibinfo {pages} {226301} (\bibinfo {year} {2024})}\BibitemShut {NoStop}%
\bibitem [{\citenamefont {Blonder}\ \emph {et~al.}(1982)\citenamefont {Blonder}, \citenamefont {Tinkham},\ and\ \citenamefont {Klapwijk}}]{blonder1982transition}%
  \BibitemOpen
  \bibfield  {author} {\bibinfo {author} {\bibfnamefont {G.}~\bibnamefont {Blonder}}, \bibinfo {author} {\bibfnamefont {m.~M.}\ \bibnamefont {Tinkham}}, \ and\ \bibinfo {author} {\bibfnamefont {T.}~\bibnamefont {Klapwijk}},\ }\href@noop {} {\bibfield  {journal} {\bibinfo  {journal} {Phys. Rev. B}\ }\textbf {\bibinfo {volume} {25}},\ \bibinfo {pages} {4515} (\bibinfo {year} {1982})}\BibitemShut {NoStop}%
\bibitem [{\citenamefont {Beenakker}(1992)}]{beenakker1992quantum}%
  \BibitemOpen
  \bibfield  {author} {\bibinfo {author} {\bibfnamefont {C.}~\bibnamefont {Beenakker}},\ }\href@noop {} {\bibfield  {journal} {\bibinfo  {journal} {Phys. Rev. B}\ }\textbf {\bibinfo {volume} {46}},\ \bibinfo {pages} {12841} (\bibinfo {year} {1992})}\BibitemShut {NoStop}%
\bibitem [{\citenamefont {Blonder}\ and\ \citenamefont {Tinkham}(1983)}]{BTK_cunb_pointcontacts}%
  \BibitemOpen
  \bibfield  {author} {\bibinfo {author} {\bibfnamefont {G.~E.}\ \bibnamefont {Blonder}}\ and\ \bibinfo {author} {\bibfnamefont {M.}~\bibnamefont {Tinkham}},\ }\href {\doibase 10.1103/PhysRevB.27.112} {\bibfield  {journal} {\bibinfo  {journal} {Phys. Rev. B}\ }\textbf {\bibinfo {volume} {27}},\ \bibinfo {pages} {112} (\bibinfo {year} {1983})}\BibitemShut {NoStop}%
\bibitem [{\citenamefont {Zaitsev}(1980)}]{Zaitsev1980}%
  \BibitemOpen
  \bibfield  {author} {\bibinfo {author} {\bibfnamefont {A.~V.}\ \bibnamefont {Zaitsev}},\ }\href@noop {} {\bibfield  {journal} {\bibinfo  {journal} {Sov. Phys. JETP}\ }\textbf {\bibinfo {volume} {51}},\ \bibinfo {pages} {111} (\bibinfo {year} {1980})},\ \bibinfo {note} {original Russian publication: Zh. Eksp. Teor. Fiz. 78, 221 (1980)}\BibitemShut {NoStop}%
\bibitem [{\citenamefont {Artemenko}\ \emph {et~al.}(1979)\citenamefont {Artemenko}, \citenamefont {Volkov},\ and\ \citenamefont {Zaitsev}}]{Artemenko1979}%
  \BibitemOpen
  \bibfield  {author} {\bibinfo {author} {\bibfnamefont {N.}~\bibnamefont {Artemenko}}, \bibinfo {author} {\bibfnamefont {A.~F.}\ \bibnamefont {Volkov}}, \ and\ \bibinfo {author} {\bibfnamefont {A.~V.}\ \bibnamefont {Zaitsev}},\ }\href@noop {} {\bibfield  {journal} {\bibinfo  {journal} {Solid State Commun.}\ }\textbf {\bibinfo {volume} {30}},\ \bibinfo {pages} {771} (\bibinfo {year} {1979})}\BibitemShut {NoStop}%
\bibitem [{\citenamefont {Octavio}\ \emph {et~al.}(1983)\citenamefont {Octavio}, \citenamefont {Tinkham}, \citenamefont {Blonder},\ and\ \citenamefont {Klapwijk}}]{octavio1983subharmonic}%
  \BibitemOpen
  \bibfield  {author} {\bibinfo {author} {\bibfnamefont {M.}~\bibnamefont {Octavio}}, \bibinfo {author} {\bibfnamefont {M.}~\bibnamefont {Tinkham}}, \bibinfo {author} {\bibfnamefont {G.}~\bibnamefont {Blonder}}, \ and\ \bibinfo {author} {\bibfnamefont {T.}~\bibnamefont {Klapwijk}},\ }\href@noop {} {\bibfield  {journal} {\bibinfo  {journal} {Phys. Rev. B}\ }\textbf {\bibinfo {volume} {27}},\ \bibinfo {pages} {6739} (\bibinfo {year} {1983})}\BibitemShut {NoStop}%
\bibitem [{\citenamefont {Gerber}\ \emph {et~al.}(2025)\citenamefont {Gerber}, \citenamefont {Ersoy}, \citenamefont {Masseroni}, \citenamefont {Niese}, \citenamefont {Laumer}, \citenamefont {Denisov}, \citenamefont {Duprez}, \citenamefont {Huang}, \citenamefont {Adam}, \citenamefont {Ostertag}, \citenamefont {Tong}, \citenamefont {Taniguchi}, \citenamefont {Watanabe}, \citenamefont {Fal'ko}, \citenamefont {Ihn}, \citenamefont {Ensslin},\ and\ \citenamefont {Knothe}}]{Jonas_QPC}%
  \BibitemOpen
  \bibfield  {author} {\bibinfo {author} {\bibfnamefont {J.~D.}\ \bibnamefont {Gerber}}, \bibinfo {author} {\bibfnamefont {E.}~\bibnamefont {Ersoy}}, \bibinfo {author} {\bibfnamefont {M.}~\bibnamefont {Masseroni}}, \bibinfo {author} {\bibfnamefont {M.}~\bibnamefont {Niese}}, \bibinfo {author} {\bibfnamefont {M.}~\bibnamefont {Laumer}}, \bibinfo {author} {\bibfnamefont {A.~O.}\ \bibnamefont {Denisov}}, \bibinfo {author} {\bibfnamefont {H.}~\bibnamefont {Duprez}}, \bibinfo {author} {\bibfnamefont {W.~W.}\ \bibnamefont {Huang}}, \bibinfo {author} {\bibfnamefont {C.}~\bibnamefont {Adam}}, \bibinfo {author} {\bibfnamefont {L.}~\bibnamefont {Ostertag}}, \bibinfo {author} {\bibfnamefont {C.}~\bibnamefont {Tong}}, \bibinfo {author} {\bibfnamefont {T.}~\bibnamefont {Taniguchi}}, \bibinfo {author} {\bibfnamefont {K.}~\bibnamefont {Watanabe}}, \bibinfo {author} {\bibfnamefont {V.~I.}\ \bibnamefont {Fal'ko}}, \bibinfo {author} {\bibfnamefont {T.}~\bibnamefont {Ihn}}, \bibinfo {author} {\bibfnamefont {K.}~\bibnamefont
  {Ensslin}}, \ and\ \bibinfo {author} {\bibfnamefont {A.}~\bibnamefont {Knothe}},\ }\href {\doibase 10.1021/acs.nanolett.5c02309} {\bibfield  {journal} {\bibinfo  {journal} {Nano Letters}\ }\textbf {\bibinfo {volume} {25}},\ \bibinfo {pages} {12480} (\bibinfo {year} {2025})}\BibitemShut {NoStop}%
\bibitem [{\citenamefont {Overweg}\ \emph {et~al.}(2018{\natexlab{b}})\citenamefont {Overweg}, \citenamefont {Knothe}, \citenamefont {Fabian}, \citenamefont {Linhart}, \citenamefont {Rickhaus}, \citenamefont {Wernli}, \citenamefont {Watanabe}, \citenamefont {Taniguchi}, \citenamefont {S\'anchez}, \citenamefont {Burgd\"orfer}, \citenamefont {Libisch}, \citenamefont {Fal'ko}, \citenamefont {Ensslin},\ and\ \citenamefont {Ihn}}]{topologically_states_QPC}%
  \BibitemOpen
  \bibfield  {author} {\bibinfo {author} {\bibfnamefont {H.}~\bibnamefont {Overweg}}, \bibinfo {author} {\bibfnamefont {A.}~\bibnamefont {Knothe}}, \bibinfo {author} {\bibfnamefont {T.}~\bibnamefont {Fabian}}, \bibinfo {author} {\bibfnamefont {L.}~\bibnamefont {Linhart}}, \bibinfo {author} {\bibfnamefont {P.}~\bibnamefont {Rickhaus}}, \bibinfo {author} {\bibfnamefont {L.}~\bibnamefont {Wernli}}, \bibinfo {author} {\bibfnamefont {K.}~\bibnamefont {Watanabe}}, \bibinfo {author} {\bibfnamefont {T.}~\bibnamefont {Taniguchi}}, \bibinfo {author} {\bibfnamefont {D.}~\bibnamefont {S\'anchez}}, \bibinfo {author} {\bibfnamefont {J.}~\bibnamefont {Burgd\"orfer}}, \bibinfo {author} {\bibfnamefont {F.}~\bibnamefont {Libisch}}, \bibinfo {author} {\bibfnamefont {V.~I.}\ \bibnamefont {Fal'ko}}, \bibinfo {author} {\bibfnamefont {K.}~\bibnamefont {Ensslin}}, \ and\ \bibinfo {author} {\bibfnamefont {T.}~\bibnamefont {Ihn}},\ }\href {\doibase 10.1103/PhysRevLett.121.257702} {\bibfield  {journal} {\bibinfo  {journal} {Phys. Rev.
  Lett.}\ }\textbf {\bibinfo {volume} {121}},\ \bibinfo {pages} {257702} (\bibinfo {year} {2018}{\natexlab{b}})}\BibitemShut {NoStop}%
\bibitem [{\citenamefont {Linder}\ and\ \citenamefont {Sudb\o{}}(2008)}]{FWM}%
  \BibitemOpen
  \bibfield  {author} {\bibinfo {author} {\bibfnamefont {J.}~\bibnamefont {Linder}}\ and\ \bibinfo {author} {\bibfnamefont {A.}~\bibnamefont {Sudb\o{}}},\ }\href {\doibase 10.1103/PhysRevB.77.064507} {\bibfield  {journal} {\bibinfo  {journal} {Phys. Rev. B}\ }\textbf {\bibinfo {volume} {77}},\ \bibinfo {pages} {064507} (\bibinfo {year} {2008})}\BibitemShut {NoStop}%
\bibitem [{\citenamefont {Delfanazari}\ \emph {et~al.}(2024)\citenamefont {Delfanazari}, \citenamefont {Li}, \citenamefont {Xiong}, \citenamefont {Ma}, \citenamefont {Puddy}, \citenamefont {Yi}, \citenamefont {Farrer}, \citenamefont {Komori}, \citenamefont {Robinson}, \citenamefont {Serra} \emph {et~al.}}]{hybrid_arrays_2024}%
  \BibitemOpen
  \bibfield  {author} {\bibinfo {author} {\bibfnamefont {K.}~\bibnamefont {Delfanazari}}, \bibinfo {author} {\bibfnamefont {J.}~\bibnamefont {Li}}, \bibinfo {author} {\bibfnamefont {Y.}~\bibnamefont {Xiong}}, \bibinfo {author} {\bibfnamefont {P.}~\bibnamefont {Ma}}, \bibinfo {author} {\bibfnamefont {R.~K.}\ \bibnamefont {Puddy}}, \bibinfo {author} {\bibfnamefont {T.}~\bibnamefont {Yi}}, \bibinfo {author} {\bibfnamefont {I.}~\bibnamefont {Farrer}}, \bibinfo {author} {\bibfnamefont {S.}~\bibnamefont {Komori}}, \bibinfo {author} {\bibfnamefont {J.~W.}\ \bibnamefont {Robinson}}, \bibinfo {author} {\bibfnamefont {L.}~\bibnamefont {Serra}},  \emph {et~al.},\ }\href@noop {} {\bibfield  {journal} {\bibinfo  {journal} {Phys. Rev. Appl.}\ }\textbf {\bibinfo {volume} {21}},\ \bibinfo {pages} {014051} (\bibinfo {year} {2024})}\BibitemShut {NoStop}%
\bibitem [{\citenamefont {Gao}\ \emph {et~al.}(2025)\citenamefont {Gao}, \citenamefont {Song}, \citenamefont {Wang}, \citenamefont {Geng}, \citenamefont {Cao}, \citenamefont {Yu}, \citenamefont {Yang}, \citenamefont {Xu}, \citenamefont {Chen}, \citenamefont {Li} \emph {et~al.}}]{nanowires_2025}%
  \BibitemOpen
  \bibfield  {author} {\bibinfo {author} {\bibfnamefont {Y.}~\bibnamefont {Gao}}, \bibinfo {author} {\bibfnamefont {W.}~\bibnamefont {Song}}, \bibinfo {author} {\bibfnamefont {Y.}~\bibnamefont {Wang}}, \bibinfo {author} {\bibfnamefont {Z.}~\bibnamefont {Geng}}, \bibinfo {author} {\bibfnamefont {Z.}~\bibnamefont {Cao}}, \bibinfo {author} {\bibfnamefont {Z.}~\bibnamefont {Yu}}, \bibinfo {author} {\bibfnamefont {S.}~\bibnamefont {Yang}}, \bibinfo {author} {\bibfnamefont {J.}~\bibnamefont {Xu}}, \bibinfo {author} {\bibfnamefont {F.}~\bibnamefont {Chen}}, \bibinfo {author} {\bibfnamefont {Z.}~\bibnamefont {Li}},  \emph {et~al.},\ }\href@noop {} {\bibfield  {journal} {\bibinfo  {journal} {Phys. Rev. Appl.}\ }\textbf {\bibinfo {volume} {23}},\ \bibinfo {pages} {L061004} (\bibinfo {year} {2025})}\BibitemShut {NoStop}%
\bibitem [{\citenamefont {Court}\ \emph {et~al.}(2007)\citenamefont {Court}, \citenamefont {Ferguson},\ and\ \citenamefont {Clark}}]{Gap_vs_d_spectroscopy}%
  \BibitemOpen
  \bibfield  {author} {\bibinfo {author} {\bibfnamefont {N.}~\bibnamefont {Court}}, \bibinfo {author} {\bibfnamefont {A.}~\bibnamefont {Ferguson}}, \ and\ \bibinfo {author} {\bibfnamefont {R.}~\bibnamefont {Clark}},\ }\href {\doibase 10.1088/0953--2048/21/01/015013} {\bibfield  {journal} {\bibinfo  {journal} {Supercond. Sci. Technol.}\ }\textbf {\bibinfo {volume} {21}},\ \bibinfo {pages} {015013} (\bibinfo {year} {2007})}\BibitemShut {NoStop}%
\bibitem [{\citenamefont {Marchegiani}\ \emph {et~al.}(2022)\citenamefont {Marchegiani}, \citenamefont {Amico},\ and\ \citenamefont {Catelani}}]{Bulk_value}%
  \BibitemOpen
  \bibfield  {author} {\bibinfo {author} {\bibfnamefont {G.}~\bibnamefont {Marchegiani}}, \bibinfo {author} {\bibfnamefont {L.}~\bibnamefont {Amico}}, \ and\ \bibinfo {author} {\bibfnamefont {G.}~\bibnamefont {Catelani}},\ }\href {\doibase 10.1103/PRXQuantum.3.040338} {\bibfield  {journal} {\bibinfo  {journal} {PRX Quantum}\ }\textbf {\bibinfo {volume} {3}},\ \bibinfo {pages} {040338} (\bibinfo {year} {2022})}\BibitemShut {NoStop}%
\bibitem [{\citenamefont {Martinis}\ \emph {et~al.}(2000)\citenamefont {Martinis}, \citenamefont {Hilton}, \citenamefont {Irwin},\ and\ \citenamefont {Wollman}}]{MARTINIS200023}%
  \BibitemOpen
  \bibfield  {author} {\bibinfo {author} {\bibfnamefont {J.~M.}\ \bibnamefont {Martinis}}, \bibinfo {author} {\bibfnamefont {G.}~\bibnamefont {Hilton}}, \bibinfo {author} {\bibfnamefont {K.}~\bibnamefont {Irwin}}, \ and\ \bibinfo {author} {\bibfnamefont {D.}~\bibnamefont {Wollman}},\ }\href {\doibase https://doi.org/10.1016/S0168--9002(99)01320--0} {\bibfield  {journal} {\bibinfo  {journal} {Nucl. Instrum. Methods Phys. Res., Sect. A}\ }\textbf {\bibinfo {volume} {444}},\ \bibinfo {pages} {23} (\bibinfo {year} {2000})}\BibitemShut {NoStop}%
\bibitem [{\citenamefont {Steffen}\ \emph {et~al.}(2023)\citenamefont {Steffen}, \citenamefont {Dutta}, \citenamefont {Wang}, \citenamefont {Li}, \citenamefont {Huang}, \citenamefont {Huang}, \citenamefont {Mathur}, \citenamefont {Wellstood},\ and\ \citenamefont {Palmer}}]{ALO_junctions}%
  \BibitemOpen
  \bibfield  {author} {\bibinfo {author} {\bibfnamefont {Z.}~\bibnamefont {Steffen}}, \bibinfo {author} {\bibfnamefont {s.}~\bibnamefont {Dutta}}, \bibinfo {author} {\bibfnamefont {H.}~\bibnamefont {Wang}}, \bibinfo {author} {\bibfnamefont {K.}~\bibnamefont {Li}}, \bibinfo {author} {\bibfnamefont {Y.}~\bibnamefont {Huang}}, \bibinfo {author} {\bibfnamefont {Y.-H.}\ \bibnamefont {Huang}}, \bibinfo {author} {\bibfnamefont {A.}~\bibnamefont {Mathur}}, \bibinfo {author} {\bibfnamefont {F.}~\bibnamefont {Wellstood}}, \ and\ \bibinfo {author} {\bibfnamefont {B.}~\bibnamefont {Palmer}},\ }\href {\doibase 10.1109/TASC.2023.3247987} {\bibfield  {journal} {\bibinfo  {journal} {IEEE Trans. Appl. Supercond.}\ }\textbf {\bibinfo {volume} {33}},\ \bibinfo {pages} {1} (\bibinfo {year} {2023})}\BibitemShut {NoStop}%
\bibitem [{\citenamefont {Choi}\ \emph {et~al.}(2010)\citenamefont {Choi}, \citenamefont {Lee},\ and\ \citenamefont {Doh}}]{YongJoo_2010}%
  \BibitemOpen
  \bibfield  {author} {\bibinfo {author} {\bibfnamefont {J.-H.}\ \bibnamefont {Choi}}, \bibinfo {author} {\bibfnamefont {H.-J.}\ \bibnamefont {Lee}}, \ and\ \bibinfo {author} {\bibfnamefont {Y.-J.}\ \bibnamefont {Doh}},\ }\href {\doibase 10.3938/jkps.57.149} {\bibfield  {journal} {\bibinfo  {journal} {J. Korean Phys. Soc.}\ }\textbf {\bibinfo {volume} {57}} (\bibinfo {year} {2010}),\ 10.3938/jkps.57.149}\BibitemShut {NoStop}%
\bibitem [{\citenamefont {Jehl}\ \emph {et~al.}(2000)\citenamefont {Jehl}, \citenamefont {Sanquer}, \citenamefont {Calemczuk},\ and\ \citenamefont {Mailly}}]{Jehl2000}%
  \BibitemOpen
  \bibfield  {author} {\bibinfo {author} {\bibfnamefont {X.}~\bibnamefont {Jehl}}, \bibinfo {author} {\bibfnamefont {M.}~\bibnamefont {Sanquer}}, \bibinfo {author} {\bibfnamefont {R.}~\bibnamefont {Calemczuk}}, \ and\ \bibinfo {author} {\bibfnamefont {D.}~\bibnamefont {Mailly}},\ }\href {\doibase 10.1038/35011012} {\bibfield  {journal} {\bibinfo  {journal} {Nature}\ }\textbf {\bibinfo {volume} {405}},\ \bibinfo {pages} {50} (\bibinfo {year} {2000})}\BibitemShut {NoStop}%
\bibitem [{\citenamefont {Kozhevnikov}\ \emph {et~al.}(2000)\citenamefont {Kozhevnikov}, \citenamefont {Schoelkopf},\ and\ \citenamefont {Prober}}]{PhysRevLett.84.3398}%
  \BibitemOpen
  \bibfield  {author} {\bibinfo {author} {\bibfnamefont {A.~A.}\ \bibnamefont {Kozhevnikov}}, \bibinfo {author} {\bibfnamefont {R.~J.}\ \bibnamefont {Schoelkopf}}, \ and\ \bibinfo {author} {\bibfnamefont {D.~E.}\ \bibnamefont {Prober}},\ }\href {\doibase 10.1103/PhysRevLett.84.3398} {\bibfield  {journal} {\bibinfo  {journal} {Phys. Rev. Lett.}\ }\textbf {\bibinfo {volume} {84}},\ \bibinfo {pages} {3398} (\bibinfo {year} {2000})}\BibitemShut {NoStop}%
\bibitem [{\citenamefont {Courtois}\ \emph {et~al.}(1999)\citenamefont {Courtois}, \citenamefont {Charlat}, \citenamefont {Gandit}, \citenamefont {Mailly},\ and\ \citenamefont {Pannetier}}]{Courtois1999}%
  \BibitemOpen
  \bibfield  {author} {\bibinfo {author} {\bibfnamefont {H.}~\bibnamefont {Courtois}}, \bibinfo {author} {\bibfnamefont {P.}~\bibnamefont {Charlat}}, \bibinfo {author} {\bibfnamefont {P.}~\bibnamefont {Gandit}}, \bibinfo {author} {\bibfnamefont {D.}~\bibnamefont {Mailly}}, \ and\ \bibinfo {author} {\bibfnamefont {B.}~\bibnamefont {Pannetier}},\ }\href {\doibase 10.1023/A:1021885617107} {\bibfield  {journal} {\bibinfo  {journal} {Journal of Low Temperature Physics}\ }\textbf {\bibinfo {volume} {116}},\ \bibinfo {pages} {187} (\bibinfo {year} {1999})}\BibitemShut {NoStop}%
\bibitem [{\citenamefont {Xiong}\ \emph {et~al.}(1993)\citenamefont {Xiong}, \citenamefont {Xiao},\ and\ \citenamefont {Laibowitz}}]{Xiong_SN_junction}%
  \BibitemOpen
  \bibfield  {author} {\bibinfo {author} {\bibfnamefont {P.}~\bibnamefont {Xiong}}, \bibinfo {author} {\bibfnamefont {G.}~\bibnamefont {Xiao}}, \ and\ \bibinfo {author} {\bibfnamefont {R.~B.}\ \bibnamefont {Laibowitz}},\ }\href {\doibase 10.1103/PhysRevLett.71.1907} {\bibfield  {journal} {\bibinfo  {journal} {Phys. Rev. Lett.}\ }\textbf {\bibinfo {volume} {71}},\ \bibinfo {pages} {1907} (\bibinfo {year} {1993})}\BibitemShut {NoStop}%
\bibitem [{\citenamefont {Jiang}\ \emph {et~al.}(2016)\citenamefont {Jiang}, \citenamefont {Zhang}, \citenamefont {Khim}, \citenamefont {Bhoi}, \citenamefont {Kim}, \citenamefont {Greene},\ and\ \citenamefont {Takeuchi}}]{jiang2016unconventional}%
  \BibitemOpen
  \bibfield  {author} {\bibinfo {author} {\bibfnamefont {Y.}~\bibnamefont {Jiang}}, \bibinfo {author} {\bibfnamefont {X.}~\bibnamefont {Zhang}}, \bibinfo {author} {\bibfnamefont {S.}~\bibnamefont {Khim}}, \bibinfo {author} {\bibfnamefont {D.}~\bibnamefont {Bhoi}}, \bibinfo {author} {\bibfnamefont {K.~H.}\ \bibnamefont {Kim}}, \bibinfo {author} {\bibfnamefont {R.~L.}\ \bibnamefont {Greene}}, \ and\ \bibinfo {author} {\bibfnamefont {I.}~\bibnamefont {Takeuchi}},\ }\href@noop {} {\bibfield  {journal} {\bibinfo  {journal} {AIP Adv.}\ }\textbf {\bibinfo {volume} {6}},\ \bibinfo {pages} {045210} (\bibinfo {year} {2016})}\BibitemShut {NoStop}%
\bibitem [{\citenamefont {Westbrook}\ and\ \citenamefont {Javan}(1999)}]{WESTBROOK_1999}%
  \BibitemOpen
  \bibfield  {author} {\bibinfo {author} {\bibfnamefont {P.~S.}\ \bibnamefont {Westbrook}}\ and\ \bibinfo {author} {\bibfnamefont {A.}~\bibnamefont {Javan}},\ }\href {\doibase 10.1103/PhysRevB.59.14606} {\bibfield  {journal} {\bibinfo  {journal} {Phys. Rev. B}\ }\textbf {\bibinfo {volume} {59}},\ \bibinfo {pages} {14606} (\bibinfo {year} {1999})}\BibitemShut {NoStop}%
\bibitem [{\citenamefont {Gifford}\ \emph {et~al.}(2016)\citenamefont {Gifford}, \citenamefont {Zhao}, \citenamefont {Li}, \citenamefont {Zhang}, \citenamefont {Kim},\ and\ \citenamefont {Chen}}]{Gifford_2016}%
  \BibitemOpen
  \bibfield  {author} {\bibinfo {author} {\bibfnamefont {J.}~\bibnamefont {Gifford}}, \bibinfo {author} {\bibfnamefont {G.}~\bibnamefont {Zhao}}, \bibinfo {author} {\bibfnamefont {B.}~\bibnamefont {Li}}, \bibinfo {author} {\bibfnamefont {J.}~\bibnamefont {Zhang}}, \bibinfo {author} {\bibfnamefont {D.}~\bibnamefont {Kim}}, \ and\ \bibinfo {author} {\bibfnamefont {T.}~\bibnamefont {Chen}},\ }\href@noop {} {\bibfield  {journal} {\bibinfo  {journal} {J. Appl. Phys.}\ }\textbf {\bibinfo {volume} {120}},\ \bibinfo {pages} {163901} (\bibinfo {year} {2016})}\BibitemShut {NoStop}%
\bibitem [{\citenamefont {Nguyen}\ \emph {et~al.}(1992)\citenamefont {Nguyen}, \citenamefont {Kroemer},\ and\ \citenamefont {Hu}}]{anomalus_MAR}%
  \BibitemOpen
  \bibfield  {author} {\bibinfo {author} {\bibfnamefont {C.}~\bibnamefont {Nguyen}}, \bibinfo {author} {\bibfnamefont {H.}~\bibnamefont {Kroemer}}, \ and\ \bibinfo {author} {\bibfnamefont {E.~L.}\ \bibnamefont {Hu}},\ }\href@noop {} {\bibfield  {journal} {\bibinfo  {journal} {Phys. Rev. Lett.}\ }\textbf {\bibinfo {volume} {69}},\ \bibinfo {pages} {2847} (\bibinfo {year} {1992})}\BibitemShut {NoStop}%
\bibitem [{\citenamefont {Du}\ \emph {et~al.}(2008{\natexlab{b}})\citenamefont {Du}, \citenamefont {Skachko},\ and\ \citenamefont {Andrei}}]{MAR_gJJ}%
  \BibitemOpen
  \bibfield  {author} {\bibinfo {author} {\bibfnamefont {X.}~\bibnamefont {Du}}, \bibinfo {author} {\bibfnamefont {I.}~\bibnamefont {Skachko}}, \ and\ \bibinfo {author} {\bibfnamefont {E.~Y.}\ \bibnamefont {Andrei}},\ }\href {\doibase 10.1103/PhysRevB.77.184507} {\bibfield  {journal} {\bibinfo  {journal} {Phys. Rev. B}\ }\textbf {\bibinfo {volume} {77}},\ \bibinfo {pages} {184507} (\bibinfo {year} {2008}{\natexlab{b}})}\BibitemShut {NoStop}%
\bibitem [{\citenamefont {Tomi}\ \emph {et~al.}(2021)\citenamefont {Tomi}, \citenamefont {Samatov}, \citenamefont {Vasenko}, \citenamefont {Laitinen}, \citenamefont {Hakonen},\ and\ \citenamefont {Golubev}}]{Tomi_2021}%
  \BibitemOpen
  \bibfield  {author} {\bibinfo {author} {\bibfnamefont {M.}~\bibnamefont {Tomi}}, \bibinfo {author} {\bibfnamefont {M.~R.}\ \bibnamefont {Samatov}}, \bibinfo {author} {\bibfnamefont {A.~S.}\ \bibnamefont {Vasenko}}, \bibinfo {author} {\bibfnamefont {A.}~\bibnamefont {Laitinen}}, \bibinfo {author} {\bibfnamefont {P.}~\bibnamefont {Hakonen}}, \ and\ \bibinfo {author} {\bibfnamefont {D.~S.}\ \bibnamefont {Golubev}},\ }\href {\doibase 10.1103/PhysRevB.104.134513} {\bibfield  {journal} {\bibinfo  {journal} {Phys. Rev. B}\ }\textbf {\bibinfo {volume} {104}},\ \bibinfo {pages} {134513} (\bibinfo {year} {2021})}\BibitemShut {NoStop}%
\bibitem [{\citenamefont {Ibabe}\ \emph {et~al.}(2023)\citenamefont {Ibabe}, \citenamefont {Gómez}, \citenamefont {Steffensen}, \citenamefont {Kanne}, \citenamefont {Nygård}, \citenamefont {Levy~Yeyati},\ and\ \citenamefont {Lee}}]{Joule_spectroscopy}%
  \BibitemOpen
  \bibfield  {author} {\bibinfo {author} {\bibfnamefont {A.}~\bibnamefont {Ibabe}}, \bibinfo {author} {\bibfnamefont {M.}~\bibnamefont {Gómez}}, \bibinfo {author} {\bibfnamefont {G.~O.}\ \bibnamefont {Steffensen}}, \bibinfo {author} {\bibfnamefont {T.}~\bibnamefont {Kanne}}, \bibinfo {author} {\bibfnamefont {J.}~\bibnamefont {Nygård}}, \bibinfo {author} {\bibfnamefont {A.}~\bibnamefont {Levy~Yeyati}}, \ and\ \bibinfo {author} {\bibfnamefont {E.~J.~H.}\ \bibnamefont {Lee}},\ }\href {\doibase 10.1038/s41467--023--38533--2} {\bibfield  {journal} {\bibinfo  {journal} {Nature Communications}\ }\textbf {\bibinfo {volume} {14}},\ \bibinfo {pages} {2873} (\bibinfo {year} {2023})}\BibitemShut {NoStop}%
\bibitem [{\citenamefont {Shpagina}\ \emph {et~al.}(2024)\citenamefont {Shpagina}, \citenamefont {Tikhonov}, \citenamefont {Ruhstorfer}, \citenamefont {Koblm\"uller},\ and\ \citenamefont {Khrapai}}]{PhysRevB.109.L140501}%
  \BibitemOpen
  \bibfield  {author} {\bibinfo {author} {\bibfnamefont {E.~V.}\ \bibnamefont {Shpagina}}, \bibinfo {author} {\bibfnamefont {E.~S.}\ \bibnamefont {Tikhonov}}, \bibinfo {author} {\bibfnamefont {D.}~\bibnamefont {Ruhstorfer}}, \bibinfo {author} {\bibfnamefont {G.}~\bibnamefont {Koblm\"uller}}, \ and\ \bibinfo {author} {\bibfnamefont {V.~S.}\ \bibnamefont {Khrapai}},\ }\href {\doibase 10.1103/PhysRevB.109.L140501} {\bibfield  {journal} {\bibinfo  {journal} {Phys. Rev. B}\ }\textbf {\bibinfo {volume} {109}},\ \bibinfo {pages} {L140501} (\bibinfo {year} {2024})}\BibitemShut {NoStop}%
\bibitem [{\citenamefont {Naidyuk}\ and\ \citenamefont {Gloos}(2018)}]{Review_PC_andreev}%
  \BibitemOpen
  \bibfield  {author} {\bibinfo {author} {\bibfnamefont {Y.~G.}\ \bibnamefont {Naidyuk}}\ and\ \bibinfo {author} {\bibfnamefont {K.}~\bibnamefont {Gloos}},\ }\href@noop {} {\bibfield  {journal} {\bibinfo  {journal} {Low Temp. Phys.}\ }\textbf {\bibinfo {volume} {44}},\ \bibinfo {pages} {257} (\bibinfo {year} {2018})}\BibitemShut {NoStop}%
\end{thebibliography}%

\end{document}